\documentclass[11pt,reqno]{amsart}
\usepackage[utf8]{inputenc}
\usepackage{braket,amsmath,amsfonts,amsthm,url,graphicx,url,xcolor,hyperref,subcaption, multicol,dsfont, comment}

\usepackage[backend=bibtex8,sorting=nty,giveninits=true,doi=false,isbn=false,url=false,maxbibnames=10]{biblatex}
\renewbibmacro{in:}{}
\newbibmacro{string+doi}[1]{\iffieldundef{doi}{#1}{\href{https://dx.doi.org/\thefield{doi}}{#1}}}
\DeclareFieldFormat{title}{\usebibmacro{string+doi}{\mkbibemph{#1}}}
\DeclareFieldFormat[article]{title}{\usebibmacro{string+doi}{\mkbibquote{#1}}}
\DeclareFieldFormat[incollection]{title}{\usebibmacro{string+doi}{\mkbibquote{#1}}}                   
\DeclareFieldFormat[inproceedings]{title}{\usebibmacro{string+doi}{\mkbibquote{#1}}}

\newtheorem{lemma}{Lemma}[section]

\newtheorem{theorem}[lemma]{Theorem}

\newtheorem{cor}[lemma]{Corollary}

\newtheorem{definition}[lemma]{Definition}

\newtheorem{fact}[lemma]{Fact}

\newtheorem{assump}[lemma]{Assumption}
\newtheorem{exam}[lemma]{Example}

\newcommand{\NN}{\mathbb{N}}

\newcommand{\E}{\mathcal{E}}
\newcommand{\F}{\mathcal{F}}

\newcommand{\Id}{\mathrm{Id}}

\newcommand{\pl}{\hspace{.1cm}}

\newcommand{\tr}{\text{tr}}

\newcommand{\polylog}{\text{polylog}}

\newcommand{\precsucc}[2]{\text{ } {}^{\prec {#2}}_{\succ {#1}} \text{ } }

\DeclareUnicodeCharacter{2009}{\,} 

\title[Relative Entropy Designs]{Shallow Approximate Unitary Designs from Relative Entropy Decay of Unstructured Random Circuits}

\author[N. Laracuente]{Nicholas Laracuente}
\address{Indiana University Bloomington. Bloomington, IN. 47404} \email[Nicholas Laracuente]{nlaracu@iu.edu}

\thanks{\hspace{-5mm} Part of this work was completed by NL as an IBM Postdoc at the University of Chicago. NL is supported by Indiana University.}

\newcommand{\mbalign}{}
\newcommand{\mbnline}{}

\begin{document}

\begin{abstract}
    Approximate unitary $k$-designs are ensembles of unitary matrices whose first $k$ statistical moments approximate the Haar (uniform) measure. Shallow, random quantum circuits with temporally and spatially structured architectures are known to yield approximate $k$-designs in depth that is logarithmic in the number of qubits. It was open whether this structure is physically necessary. Shown here is shallow relative entropy convergence for standard notions of less structured, time-periodic circuits, implying additive-error approximate k-designs in nearly logarithmic depth (up to sub-logarithmic factors). These architectures include 1-dimensional brickwork and gates placed randomly according to a connected, bounded-degree graph. Relative entropy recombination methods derived herein are potentially of independent interest.
\end{abstract}

\maketitle

\section{Introduction}
In quantum information and computing, random unitaries are often sought for their use in coding theory \cite{dupuis_decoupling_2010, hayden_decoupling_2008}, cryptographic properties \cite{ji_pseudorandom_2018}, analogies to fundamental physics \cite{aaronson_complexity_2016, piroli_random_2020}, state learning \cite{schuster_random_2025, aaronson_shadow_2018, helsen_thrifty_2023}, and many other applications \cite{mele_introduction_2024}. Ensembles of uniformly distributed random unitaries are thought rare and difficult to construct, however, because most unitaries on $n$ qubits would require exponentially many elementary gates to approximate as quantum circuits. Approximate unitary $k$-designs resemble the uniform distribution in their first $k$ statistical moments. For a probability measure $\mu$ on the unitary group $U(d)$, we denote the weighted $k$-fold twirl
\begin{align} \label{eq:twirl}
\Phi_{\mu, k}(\rho) := \int U^{\otimes k} \rho U^{\otimes k \dagger} d \mu (U) \end{align}
for every input state $\rho$ on a system of dimension $d^k$. By $\Phi_{\text{Haar}, k}$ we denote such a construction with respect to the Haar (uniform) measure on $U(d)$. A measure $\mu$ is an \textit{additive-error approximate $k$-design} if its induced $k$-fold twirl is close to $\Phi_{\text{Haar}, k}$ in diamond norm (a standard measure of quantum densities' distinguishability). For a stronger notion, $\mu$ is a \textit{relative or multiplicative-error approximate $k$-design} if its induced twirl can be rewritten as a convex combination involving $\Phi_{\text{Haar}, k}$ and vice versa. See Section \ref{sec:background} for precise mathematical definitions.

The study of unitary k-designs has a long history \cite{ambainis_quantum_2007, dankert_exact_2009, gross_hypercontractivity_1975, harrow_random_2009, brandao_local_2016, harrow_approximate_2023-1, hunter-jones_unitary_2019, chen_efficient_2024, metger_simple_2024, chen_incompressibility_2024, laracuente_approximate_2024, schuster_random_2025}. Although multiplicative error is stronger, additive error is commonly studied. One line of work has sought primarily to discover efficient design constructions for coding \cite{liu_approximate_2026}, learning \cite{schuster_random_2025}, and other applications. On $n$ qubits of constant local dimension, structured random circuits can form $\epsilon$-approximate multiplicative error $k$-designs in depth $O(k \mathrm{polylog} k \log (n/\epsilon))$ \cite{schuster_random_2025}. These constructions work even in one-dimensional nearest-neighbor connectivity, defying the intuition that information would have to spread across a chain of qubits. With extra qubits and connectivity, specific algorithms achieve sub-logarithmic depth designs \cite{cui_unitary_2025, zhang_designs_2025}. The $O(\log n)$ is however the shallowest possible depth for additive-error designs via circuits composed of 2-local Haar random gates with no extra qubits \cite{dalzell_random_2022, deshpande_tight_2022}.

\begin{figure}[h!] \centering
	\begin{subfigure}[b]{0.45\textwidth}
		\includegraphics[width=0.98\textwidth]{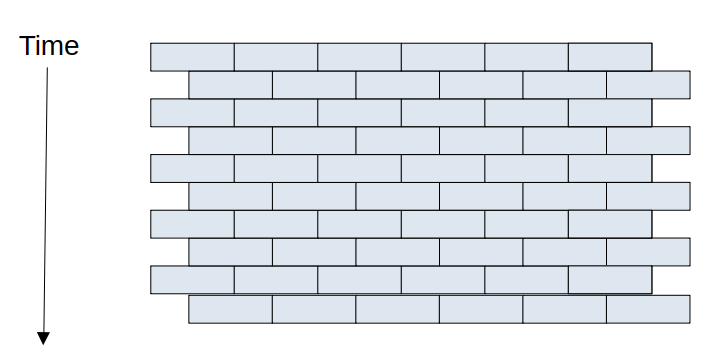}
		\caption{In a standard, one-dimensional `brickwork' architecture, each alternating layer pairs neighboring qudits and applies a random 2-qudit gate to each pair.}
	\end{subfigure}
	\hspace{5mm}
	\begin{subfigure}[b]{0.45\textwidth}
		\includegraphics[width=0.98\textwidth]{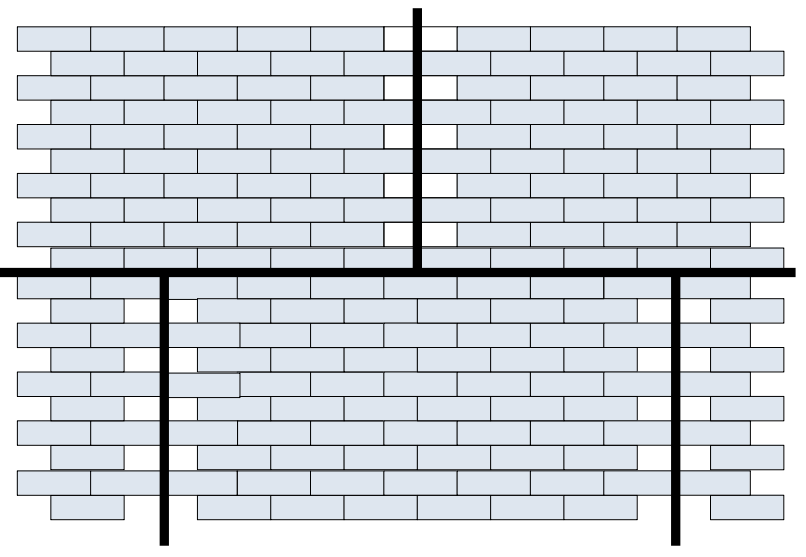}
		\caption{In the architectures of \cite{schuster_random_2025, laracuente_approximate_2024}, the quantum circuit must be split into large ``chunks'' each of which individually may apply brickwork.}
	\end{subfigure}
    \caption{}
    \label{fig:brickvschunk}
\end{figure}

A related but distinct question is how quickly unstructured or more seemingly natural random circuits converge to $k$-designs. Here, ``unstructured'' refers to architectures that are time-periodic with $O(1)$ period (the same process is repeated every few layers independently of $k$) and spatially organized in ways that do not depend on $k$. One canonical, unstructured architecture applies 2-qudit gates between randomly chosen pairs at each step \cite{harrow_random_2009, dalzell_random_2022}. Another canonical and commonly studied ``unstructured'' architecture in one spatial dimension is known as (1-dimensional) brickwork \cite{brandao_local_2016, bertini_nonequilibrium_2026}. Brickwork is an architecture on $n$ qudits with nearest neighbor connectivity, in that qudit 1 connects to 2, 2 to 3, and so on until $n-1$ and $n$. A layer of brickwork applies parallel, random, 2-qudit unitaries in alternating layers. The first layer applies random 2-qudit unitaries between qudits 1 and 2, 3 and 4, and so on. The second applies random unitaries between qudits 2 and 3, 4 and 5, etc. The state of the art for brickwork and for all-to-all random circuits set by \cite{chen_efficient_2024} is that roughly $O(n k \polylog k \log (1/\epsilon))$ parallel layers suffice for multiplicative error $\epsilon$. The same $O(n k \polylog k \log (1/\epsilon))$ was also the best shown for \emph{additive} error in 1-dimensional connectivity, although additive error admits connectivity-dependent improvements \cite{harrow_approximate_2023-1}. An advantage of unstructured schemes is that they achieve progressively higher values of $k$ as additional, architecturally identical layers are added, whereas the schemes of \cite{schuster_random_2025, laracuente_approximate_2024} require $k$-dependent architectural choices starting at the first layer (see Figure \ref{fig:brickvschunk}). Unstructured random circuits are therefore preferred as modeling stationary quantum dynamics that accrue randomness and complexity \cite{brandao_models_2021, chen_incompressibility_2024} over time.

Motivating the current work, \emph{the structured schemes of \cite{schuster_random_2025, laracuente_approximate_2024} achieve design depths logarithmic in $n$ even in minimal connectivity, while the state of the art for brickwork and random gate placements is generally linear or geometry-dependent polynomial depth}. An open question noted in \cite{schuster_random_2025} and \cite{laracuente_approximate_2024} is whether this discrepancy reflects a physical difference or a gap in proof techniques. It was shown in \cite{belkin_absence_2025} that adding gates to an architecture can slow convergence to a $k$-design, suggesting that to `de-structure' shallow $k$-designs is non-trivial.

This work initiates studying $k$-design convergence in relative entropy \cite{rouze_logarithmic_2024}. First, recall the Umegaki relative entropy defined for density matrices $\rho$ and $\omega$ on the same space by
\begin{equation}
    D(\rho \| \omega) := \tr(\rho(\log \rho - \log \omega)) \pl.
\end{equation}
The logarithm base is unimportant for most inequalities we consider (as they involve ratios), but on a system of $n$ qudits of local dimension $q$, we by default take it base $q$. The relative entropy is sometimes known as the ``mother of all entropies'' as it underlies many information-theoretic quantities, including mutual information, coherent information, and communication capacities, and resource measures \cite{seshadreesan_renyi_2015, wilde_quantum_2017}.
\begin{definition}
We call a unitary measure $\mu$ a \textbf{$\lambda$-converging entropic $k$-design} if
\begin{equation}
    D((\Phi_{\mu,k} \otimes \Id) (\rho) \| (\Phi_{\mathrm{Haar}, k} \otimes \Id) (\rho))
        \leq (1-\lambda) D(\rho\| (\Phi_{\mathrm{Haar}, k} \otimes \Id) (\rho))
\end{equation}
for every input density $\rho$ and every possible extension by the identity channel $\Id$ on a finite-dimensional auxiliary system.
\end{definition}
Entropic convergence is a case of a historically studied inequality called strong data processing (see Section \ref{sec:background} and \cite{gao_complete_2022, rouze_logarithmic_2024}). It is known that
\begin{equation} \label{eq:entropy-as-relative}
    D(\Phi_{\mu,k}(\rho) \| \Phi_{\mathrm{Haar}, k}(\rho)) = H(\Phi_{\mathrm{Haar}, k}(\rho)) - H(\Phi_{\mu,k}(\rho)) \pl,
\end{equation}
where $H(\rho) = - \tr(\rho \log \rho)$ denotes the von Neumann entropy. Via Pinsker's inequality, if $\Phi_{\mu,k}$ has $\lambda$-convergence, then any measure $\nu$ such that $\Phi_{\nu, k} = \Phi_{\mu, k}^{t}$ is an $\epsilon$-approximate additive error design when $t \geq \ln (4 n k \ln (q) / \epsilon^2) / \lambda$. Hence, entropic convergence is comparable to additive though weaker than multiplicative error.

This work's main results (Theorems \ref{thm:parallel-2layer} and \ref{thm:random-graph}) show that many unstructured random circuit ensembles converge entropically in depths \textit{almost independent of $n$}. These lead to additive error convergence in depths that are nearly optimal (logarithmic) in $n$ up to exponentials of iterated-log corrections. Furthermore, the relative entropy results are divisible, meaning that convergence occurs layer-by-layer regardless of intervening unitaries. Divisibility allows one to mix-and-match architectures and automatically extends results to a wider variety of settings (see Section \ref{sec:generalize}).

\section{Background} \label{sec:background}
The iterated logarithm $\log^*(x)$ is the number of times that the natural logarithm must be applied to $x$ before the result is less than 1. It is known to be extremely slow-growing. For example, for any fixed $r \in \NN$ and large enough $n \in \NN$, $\log^*(n)$ is smaller than $\log^{(r)}(n)$ (applying the logarithm $r$ times to $n$). Therefore, for any $a > 1$, $a^{\log^*n} \leq \log^{(r)}(n)$ for sufficiently large $n$.


A (trace-symmetric) conditional expectation $\E$ is an idempotent quantum channel that is self-adjoint with respect to the Hilbert-Schmidt inner product: $\tr(\E(X) Y) = \tr(X \E(Y))$ for every $X$ and $Y$ in the space of possible input operators. Conditional expectations are studied in operator algebras as restriction maps to von Neumann algebras \cite{von_neumann_rings_1949}.
When $\Phi \circ \E = \E \circ \Phi = \E \circ \E = \E$ for channels $\Phi$ and $\E$, we say that $\E$ is a fixed point projection of $\Phi$. The multiplicative domain projection $\E$ of a unital channel $\Phi$ projects to a ``decoherence-free'' subspace on which the channel acts invertibly \cite{kribs_quantum_2006, choi_multiplicative_2009,johnston_generalized_2011}. Note that these are subspaces of operators, not Hilbert spaces - the projection channels considered here are not vector or matrix projections, but idempotent channels. In cases considered herein, the multiplicative domain will generally be a fixed point subspace, on which the channel acts trivially. The fixed point projections studied herein, particularly $\Phi_{\mathrm{Haar}, k}$, are trace-symmetric conditional expectations. Recalling two standard notions of approximate $k$-designs, $\mu$ is an $\epsilon$-approximate \dots
\begin{itemize}
    \item \dots \textbf{additive $k$-design} if $\| \Phi_{\mu, k}  - \Phi_{\text{Haar}, k} \|_{\Diamond} \leq \epsilon$. The diamond norm for a superoperator $\Phi$ is given by
    $
        \| \Phi \|_{\Diamond} = \max_{X \neq 0} {\|(\Phi \otimes \Id^B)(X)\|_1}/{\|X\|_1} \pl,
    $
    where $\|X\|_1 := \tr(|X|)$ is the trace norm, and $B$ is an auxiliary system of the same dimension as the input. 
    \item \dots \textbf{multiplicative $k$-design} if $(1-\epsilon) \Phi_{\text{Haar}, k} 
    \prec \Phi_{\mu, k} \prec (1+\epsilon) \Phi_{\text{Haar}, k}$, where for maps $\Phi$ and $\Psi$, $\Phi \succ (1-\epsilon) \Psi$ if $\Phi - (1-\epsilon) \Psi$ is completely positive, and ``$\prec$'' is defined conversely. Multiplicative comparability is essentially stronger than additive: if $\Phi$ is $\epsilon$-multiplicatively comparable to $\Psi$, then they are within $2 \epsilon$ diamond norm distance.
\end{itemize}

The relative entropy is always non-increasing when applying the same channel to both arguments, an inequality known as `data processing.' Following conventions of prior works \cite{kastoryano_quantum_2013, bardet_estimating_2017, bardet_hypercontractivity_2022, gao_fisher_2020, gao_complete_2022}, a quantum channel $\Phi$ with fixed point projection $\E$ admits a \textbf{strong data processing inequality with constant $\lambda$ ($\lambda$-SDPI)} if
\begin{equation}
    D(\Phi(\rho) \| \Phi \circ \E(\rho)) \leq (1 - \lambda) D(\rho \| \E(\rho))
\end{equation}
for every input density $\rho$. SDPI is \textbf{`complete' ($\lambda$-CSDPI)} if the same inequality holds with the same value of $\lambda$ when $\Phi$ and $\E$ are respectively extended to $\Phi \otimes \Id$ and $\E \otimes \Id$, where $\Id$ is the identity map on an auxiliary system of arbitrarily large dimension. The assumption that a channel has $\lambda$-(C)SDPI implies that its input and output spaces are the same. Pinsker's inequality converts from relative entropy to additive error: for densities $\rho$ and $\sigma$ on the same space, when the relative entropy is defined with the natural logarithm, $\| \rho - \sigma \|_1^2 \leq  2 D(\rho \| \sigma)$.

\section{1-D Brickwork and Related Architectures} \label{sec:parallel}
In a fixed architecture, the location and time of each local sub-process or gate is given. 1-D brickwork is a common example, as are others studied in \cite{harrow_approximate_2023-1}. More formally, on $n$ qudits, an $s$-local, $\ell$-layer parallel architecture is a list of sets $(L_t)_{t=1}^\ell$, where the elements of each $L_t$ are disjoint sets of at most $s$ elements of $\{1, \dots, n\}$. Each $L_t$ defines a parallel layer, and each of its constituents defines the qudits to which a local twirl or random gate are applied. We will assume that the hypergraph in which edges correspond to gate placements is connected - otherwise, the circuit should be treated as multiple circuits applied to independent systems. One may order the individual gates in any way that is compatible with the causal order of supports and layers, obtaining a sequence of gate placements.

To analyze fixed architectures, a new relative entropy decay composition Theorem is derived as Theorem \ref{thm:compose-2stream}. The main idea of Theorem \ref{thm:compose-2stream} is that given a composition of quantum channels, one may partition the channels into contiguous subsequences and derive a CSDPI constant for the entire composition in terms of subsequences' CSDPI rates and the entropic gluing coefficients of overlapping, adjacent subsequence pairs. As Theorem \ref{thm:compose-2stream} is more technical, it and its proof are relegated to Section \ref{sec:entropy}. To illustrate its motivating consequence:
\begin{cor}[1-D Brickwork] \label{cor:brick1d}
    The one-dimensional brickwork architecture with Haar-random 2-qubit gates is a $\lambda$-converging entropic $k$-design with $\lambda = 1 / O(5^{\log^*{(2 n)}} \times k \log k)$, whether on open or periodic  (for even $n$) boundary conditions. It therefore achieves an $\epsilon$-approximate additive error $k$-design in $O(5^{\log^*n}\log (n k/\epsilon) \times k \log k )$ layers.
\end{cor}
Corollary \ref{cor:brick1d} is proven as Corollary \ref{cor:brick1d-supp} in the Supplemental Material.

More generally, we call a sequence of quantum gates connected if the hypergraph in which edges define gate placements and nodes define qudits is connected. We say that an ordering of gates in an architecture is $(c,\Delta)$-connected if for fixed $c > \Delta \geq 0$, there exists a partitioning of the sequence into contiguous subsequences of length between $c$ and $c + \Delta$ (except the first, which may be shorter) such that every such subsequence is connected, and every concatenation of contiguous subsequences is connected. Requiring each gate to overlap its predecessor would be too strict even for some forms of 1-D brickwork. Instead, $(c, \Delta)$-connectedness tolerates local deviations from such a criterion. The notion of $(c,\Delta)$-connectedness generalizes the 1-D brickwork result.

\begin{figure}[h!] \centering
    \includegraphics[width=0.50\textwidth]{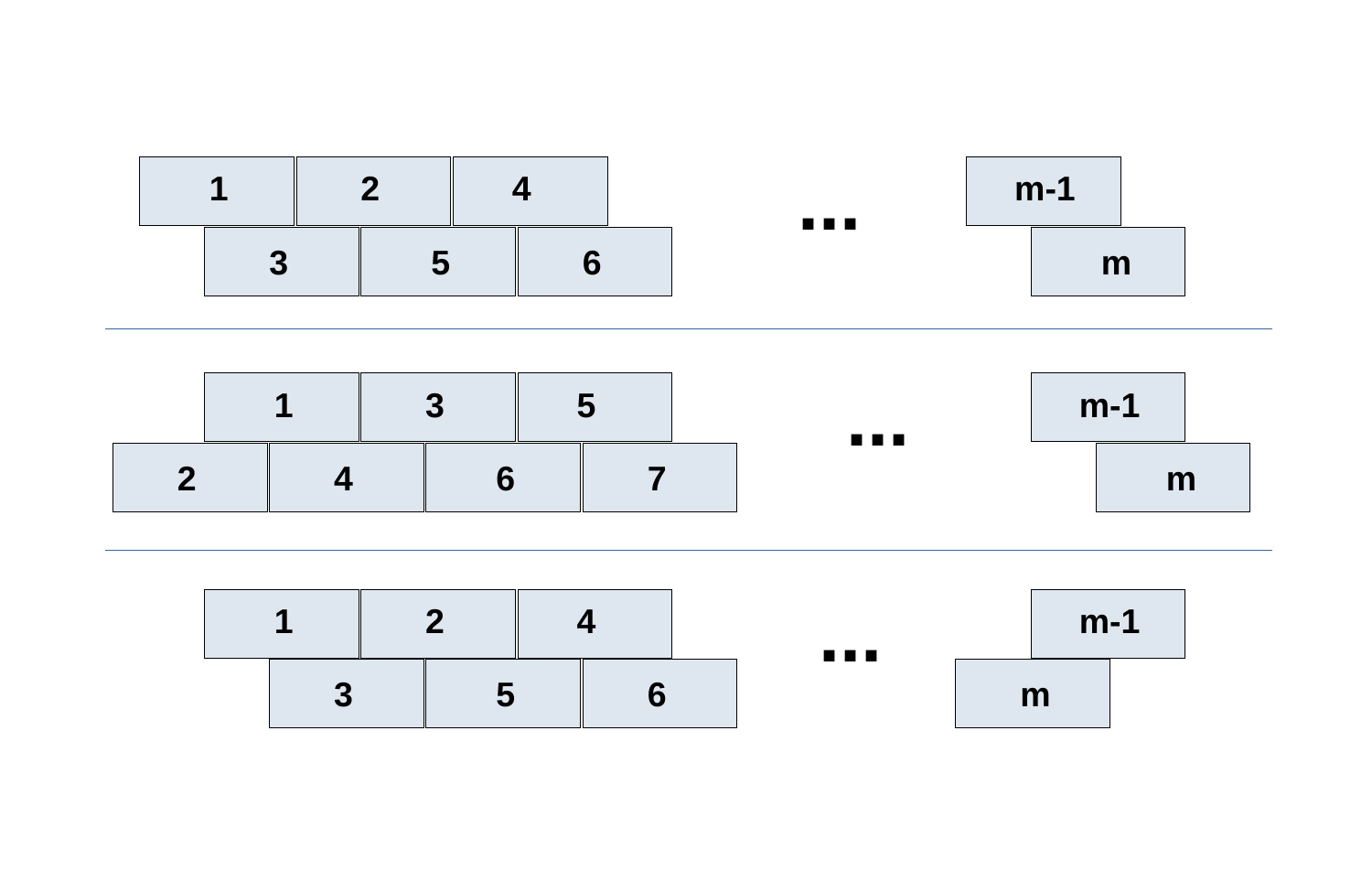}
    \caption{Illustration of orderings for brickwork with various forms of open boundary conditions.}
    \label{fig:brickwork-order}
\end{figure}

\begin{theorem} \label{thm:parallel-2layer}
    Let $\mu$ be a unitary measure on $n$ qudits of local dimension $q$. Assume $\mu$ is induced by an $s$-local, $\ell$-layer parallel architecture, which admits a $(c, \Delta)$-connected sequence with $c \geq c_*(k, \ell, q)$ for some $c_*(k, \ell, q) = O(\log k)$ for fixed $\ell$ and $q$. Assume that each subsequence or concatenated pair of adjacent subsequences has $\lambda_0$-CSDPI toward the Haar twirl on contained qudits. Then $\Phi_{\mu, k}$ is a $\lambda$-converging entropic $k$-design with $\lambda \geq \lambda_0 / 5^{\log^*(\ell n)}$. When local gates are Haar random, and $s = 2$, $\lambda_0 = 1/O(\mathrm{poly}(k))$ for fixed $q$ and $\ell$. When additionally $q=4$, $\lambda_0 = 1 / O(c \times k)$ for fixed $q$ and $\ell$.
\end{theorem}
A proof and concrete conditions are given in the Supplemental Material as Theorem \ref{thm:parallel-2layer-tech}. The main proof idea is to recursively build up CSDPI constants, at each level composing subsequences into exponentially longer concatenations. As illustrated in Supplemental Material Section \ref{sec:lattice}, the recombination methods of Theorem \ref{thm:parallel-2layer} extend beyond 1-D brickwork to higher spatial dimensions, although the formulation of these results is more technical and depends on a conjectured `base' estimate for small subsystems.

\section{Random Gates on Graphs}
Random circuits on random graphs are a common model of highly unstructured architecture. One may think of a graph as defining on which qudit pairs gates occur and a random circuit as applying a random gate to a uniformly random edge at each step. Assume $G$ is a simple, undirected graph with $n$ nodes. Let each node label correspond to a $q$-dimensional qudit in a quantum system. Define $\Phi_{G, k}$ as the channel that selects an edge of $G$ uniformly at random and applies a random, 2-qudit $k$-fold gate or twirl. To analyze this graph model, we prove a simpler analog of Theorem \ref{thm:compose-2stream} that is closer to the observations in \cite{laracuente_quasi-factorization_2022}.
\begin{lemma} \label{lem:decay-merge}
    Let $\Phi = \sum_{i=1}^\ell p_i \Phi_i$ be a quantum channel with fixed point projection $\E$ for which each $\Phi_i$ has $\lambda_i$-CSDPI toward fixed point projection $\E_i$, $\E_i \Phi_i = \Phi_i \E_i = \E_i$, $\E_i \E = \E \E_i = \E$, and $\sum_{i=1}^\ell \alpha_i D(\rho \| \E_i(\rho)) \geq D(\rho \| \E(\rho))$ (as in Assumption \ref{assump:quasifac}). Then $\Phi$ has $\min_i p_i \lambda_i / \alpha_i$-CSDPI.
\end{lemma}
The essentially algebraic proof appears as Lemma \ref{lem:decay-merge-supp} in the Supplemental Material. Before deriving shallow design convergence, we first seek $1/O(\mathrm{poly}(n))$-converging entropic $k$-designs. Although to the author's knowledge such a result does not specifically appear in the existing literature for general graphs, it is long expected. For this work it is shown explicitly in the Supplemental Material as \ref{lem:small-random}. Our shallow designs will build up via quasi-factorization, applying the $1/O(\mathrm{poly}(n))$-convergence estimates to small subgraphs.
\begin{theorem} \label{thm:random-graph}
Assume $G$ is a connected $n$-node graph of degree at most $c$. Then $\Phi_{G,k}$ is a $\lambda_0^G/(c n \exp(O(\log^*n)))$-converging entropic $k$-design, where fixing a spanning tree, $\lambda_0^G$ is the minimum CSDPI constant over all connected, max-degree $c$ subtrees having between $2$ and $n_0$ vertices, where $n_0 = O(\log k)$ ignoring the $c$ and $q$ dependencies. Let $\lambda_{\mathrm{edge}}$ denote the minimum CSDPI constant of any edge's random gate toward the local Haar twirl, which is 1 for local Haar twirls. In general, $\lambda_0^G = \lambda_{\mathrm{edge}} / O(\mathrm{poly}(k))$, ignoring $q$ and $c$ dependence. When $q=4$, $\lambda_0^G = \lambda_{\mathrm{edge}} / O(k (\log k)^2)$ ignoring $c$ dependence. When $q=2$ for a cycle or path graph,
\begin{equation} \label{eq:cycle-path}
    \lambda_{0}^G \geq \lambda_{\mathrm{edge}} / ( 2 n_0 c \lceil (2 n_0 k + \log_2(24)) / \log_2(1/(1-2^{-53})) \rceil) \pl.
\end{equation}
\end{theorem}
Theorem \ref{thm:random-graph} is proven as Theorem \ref{thm:random-graph-tech} in the Supplemental Material with concrete conditions on $n_0$. It is expected though not formally shown in the literature that $\lambda_0^G = \lambda_{\mathrm{edge}} / O(k \mathrm{polylog}(k))$ in general. Like Theorem \ref{thm:parallel-2layer}, the proof of Theorem \ref{thm:random-graph-tech} applies Lemma \ref{lem:decay-merge} recursively, building up exponentially larger graphs at each step. In contrast to the parallel circuits of Section \ref{sec:parallel}, in which each layer applies $O(n)$ gates, $\Phi_{G,k}$ applies one gate per step. Therefore, one expects to need $O(n)$ steps to achieve the analog of a single parallel layer. As in \cite{dalzell_random_2022}, the expected depth of a random gate architecture when gates are parallelized whenever possible is $O(1/n)$ times the number of steps. Theorem \ref{thm:random-graph} therefore yields additive error $k$-designs in $O(\exp(O(\log^*n)) \times \log (n/\epsilon))$ expected depth (ignoring dependence on $c, k$, and $q$). This convergence depth holds even when the graph is highly bottlenecked or has large diameter.

By convexity of relative entropy, Theorem \ref{thm:random-graph} extends to graphs for which the random gate placements can be written as convex combinations over bounded-degree, connected graphs. For example, the complete graph obeys the same CSDPI bound as does an $n$-qudit path graph - this holds because choosing gates via the complete graph is equivalent to averaging over every Hamiltonian path, which has maximum degree 2. More broadly, one may take arbitrary convex combinations of architectures, and as long as those admitting a particular CSDPI bound are sufficiently probable, the whole channel has that CSDPI bound. Theorem \ref{thm:random-graph} also extends by convexity to weighted graphs when the weights (or a connected subgraph thereof) are uniformly lower-bounded, and the decay constant is re-scaled by the probability attributable to the contained uniform graph. There are nonetheless limitations. For example, the ``lollipop graph \cite[Appendix A.C]{belkin_absence_2025}'' requires on the order of $n^2$ sequential steps to converge in additive error - that graph would also yield slower relative entropy decay than the graphs studied here.



\subsection{Emergent Designs from Interspersed Randomness} \label{sec:generalize}
Though many of the previous works noted in this paper's introduction construct circuits with the intent to produce a design, another important question is how often designs form in natural or uncontrolled settings. Unintended $2$-designs are known to cause barren plateaus in quantum optimization \cite{holmes_connecting_2022}. As relative entropy decay occurs layer-by-layer, inserting other unitary layers does not forestall entropic convergence to a $k$-design. Random circuits that yield relative entropy convergence can tolerate irregularities, inhomogeneities, and intervening processes.
\begin{exam}[Informal] \label{exam:insert} \normalfont
    Consider a quantum process that applies arbitrary unitaries over time while injecting spurious, random 2-qudit interactions at a rate $r$, which might depend on $n$. For small, fixed $k$, such a process forms an $\epsilon$-approximate additive-error $k$-design in time
    \begin{equation}
    O(n \log (n/\epsilon) \exp(O(\log^*n)) / r) \pl,
    \end{equation}
    suppressing the $k$-asymptotics and local interaction structure. In particular, a deterministic quantum circuit in which each pair of connected qudits on a bounded-degree graph has a small probability to independently undergo such a spurious gate in a given time acts in the rare-event continuous-time limit as a Poisson process that applies random gates to randomly chosen pairs. Hence, $r \sim n$ in this case, and it forms an $\epsilon$-approximate $k$-design in time $O(\log (n/\epsilon) \exp(O(\log^*n)))$.
\end{exam}
Example \ref{exam:insert} does not follow from structured ensembles or even from knowing that log-depth brickwork or that $O(n \log n)$ randomly placed gates converge to a design, because these notions do not address convergence of the individual layers. Example \ref{exam:insert} suggests a constraint on quantum circuits with imperfect controls: even if just a few random interactions occur per-qubit, per-layer, low-order statistical moments quickly resemble Haar randomness.
Prior works \cite{harrow_approximate_2023-1} asked if random circuits obey ``censoring'' inequalities: does adding gates ever slow convergence toward a $k$-design? A more recent work \cite{belkin_absence_2025} showed examples in which adding gates indeed slows convergence in additive and multiplicative error. Hence, that inserting gates between layers never slows entropy convergence is a non-trivial property.



\section{Relative Entropy Decay Recombination} \label{sec:entropy}
In this Section, we show some general bounds on how to compose relative entropy decay for compositions of channels. This section's results are used in Section \ref{sec:parallel}. They are also potentially of independent interest. For the two-sided multiplicative error comparison $(1+\delta) \Psi \succ \Phi \succ (1-\epsilon) \Psi$, we write $\Phi \precsucc{\epsilon}{\delta} \Psi$ as shorthand (or $\precsucc{\epsilon}{}$ if $\delta = \epsilon$). As shorthand, when $\E$ is a quantum channel and $X$ an expression that evaluates as a density matrix, we denote $D(X \| \E('')) := D(X \| \E(X))$. For example, $D(\Phi(\rho) \| \E('')) = D(\Phi(\rho) \| \E(\Phi(\rho)))$ for channel $\Phi$. By $\overset{\leftarrow}{\prod_{i=1}^n}$ we denote the product ordered from right to left, as is common in composing maps. An extremely useful equality is a chain rule or Pythagorean identity of relative entropy (see \cite[Theorem 1.13]{ohya_quantum_1993} or \cite[Lemma III.1]{laracuente_quasi-factorization_2022} for a more setting-specific restatement and proof):
\begin{equation} \label{eq:chain-rule}
    D(\rho \| \E(\omega)) = D(\rho \| \E(\rho)) + D(\E(\rho) \| \E(\omega))
\end{equation}
when $\E$ is a conditional expectation, and $\rho$ and $\omega$ are arbitrary states. As a useful consequence:
\begin{equation} \label{eq:chain-iter}
    D(\rho \|\omega) + D(\rho \| \E(\rho))
        \geq D(\E(\rho) \| \E(\omega)) + D(\rho \| \E(\rho))
    = D(\rho \| \E(\omega)) \pl.
\end{equation}
\subsection{The Entropic Gluing Assumption}
The following recently proven conditions imply efficient entropy recombination.
\begin{assump}[Complete $\alpha$-Entropic Gluing] \label{assump:glue} \normalfont
    Let $\E_A$ and $\E_B$ be conditional expectations and $\E_{A \cap B}$ project to their intersection algebra.
    Then $\alpha D(\rho \| \E_A \E_B (\rho)) \geq D(\rho \| \E_{A \cap B}(\rho))$, and the comparison holds under arbitrary tensor extensions $\E_X \rightarrow \E_X \otimes \Id$ for each $X \in \{A,B, A \cap B\}$.
\end{assump}
A looser analog of entropic gluing is quasi-factorization or approximate tensorization of relative entropy \cite{cesi_quasi-factorization_2001, gao_complete_2022, laracuente_quasi-factorization_2022, bardet_approximate_2022}:
\begin{assump}[Complete $\alpha$-Quasi-factorization] \label{assump:quasifac} \normalfont
    Let $\E_1, \dots, \E_\ell$ be conditional expectations and $\E$ project to their intersection algebra.
    Then for $(\alpha_j > 0)_{j=1}^\ell$, $\sum_j \alpha_j D(\rho \| \E_j(\rho)) \geq D(\rho \| \E(\rho))$, and the comparison holds under tensor extensions $\E_j \rightarrow \E_j \otimes \Id$ and $\E \rightarrow \E \otimes \Id$. We say that $\E_1, \dots, \E_\ell$ $\alpha$-quasi-factorize if quasi-factorization holds when $\alpha_j = \alpha$ for each $j$.
\end{assump}
A multiplicative error analog of Assumption \ref{assump:glue} was shown to hold as the `Gluing Lemma,' \cite[Lemmas 3 and 8]{schuster_random_2025} (and with a weaker $k$-dependence in \cite{laracuente_approximate_2024}): when $\E_A$ and $\E_B$ are each uniform $k$-fold twirls over the unitary group as in Equation \eqref{eq:twirl} on respective subsystems sharing $m$ qudits, $(1- 2 k^2 / q^m) \E_{A \cap B} \prec \E_A \E_B \prec (1 + 2 k^2 / q^m) \E_{A \cap B}$ as long as $q^m > 2 k^2$. Together with Equation \eqref{eq:quasifac} in the Supplemental Material, the Gluing Lemma implies:
\begin{fact} \label{fact:k-design-glue} \normalfont
    When $\E$ and $\E'$ are uniform $k$-fold unitary twirls as in Equation \eqref{eq:twirl} on respective subsystems sharing $m$ qudits, and $q^m > 24 k^2$, Assumption \ref{assump:glue} holds with $\alpha = (1 - 24 k^2 / q^m)^{-1}$.

    When $(\E_j)_{j=1}^\ell$ are uniform $k$-fold unitary twirls on respective subsystems such that consecutive subsystems share at least $m$ qudits, and $q^m > 24 k^2$, Assumption \ref{assump:quasifac} holds with $\alpha = (1 - 24 k^2 / q^m)^{1 - \ell}$.
\end{fact}
Other regimes in which `gluing' holds or doesn't were studied in \cite{grevink_will_2025} and \cite{west_nogo_2025}. We state `entropic gluing' as an assumption rather than quote the specific Gluing Lemma for potential applicability in other settings, such as those more traditional for relative entropy decay \cite{rouze_logarithmic_2024}.
\subsection{Entropy Recombination Results}
\begin{lemma}[Entropy Decay Decomposition] \label{lem:chainmin1}
Let $\Psi_1, \dots, \Psi_\ell$ be a family of channels such that for each $t \in 1 \dots \ell$, $\Psi_t$ has $\lambda_t$-SDPI toward and commutes with multiplicative domain conditional expectation $\F_t$. Also assume $\F$ is a conditional expectation for which $\F \F_t = \F_t \F = \F$, and $\F$ commutes with $\Psi_t$ for each $t = 1 \dots \ell$. Then
\begin{equation}
    \begin{split}
         D(\rho \| \F(\rho)) - D(\Psi_\ell \dots \Psi_1 (\rho) \| \F (''))
            \mbnline \mbalign \geq \sum_{t=1}^\ell \lambda_t D( \Psi_{t-1} \dots \Psi_1(\rho) \| \F_t ('')) \pl,
    \end{split}
\end{equation}
where $\Psi_0 \dots \Psi_1 := \Id$.
\end{lemma}
Lemma \ref{lem:chainmin1} is proven as the more general Lemma \ref{lem:chainmin1-supp} in the Supplemental Material. The common setting of Theorem \ref{thm:parallel-2layer} and Lemma \ref{lem:chainmin-strong} is introduced here to avoid repeating assumptions. Let $(\Phi_j)_{j=1}^m$ be a given family of quantum channels on the same space. For each $j = 1 \dots m$, by $\E_j$ denote the multiplicative domain conditional expectation of $\Phi_j$. For $a \leq b$ denote $\Phi[a,b] := \Phi[b] \circ \dots \circ \Phi[a]$, replacing channels with indices outside $1 \dots m$ by the identity, and by $\E[a,b]$ denote the multiplicative domain conditional expectation of $\Phi[a,b]$. Let $J \in \NN$ and $a_0, \dots, a_{J+1}$ be a strictly increasing sequence for which $a_0 = 0$, and $a_{J+1} = m+1$. The channels are assumed to obey the following conditions:
\begin{enumerate}
    \item $\Phi_t \E_t = \E_t \Phi_t = \E_t$ for each $t = 1 \dots m$;
    \item $\E[a',b'] \E[a,b] = \E[a,b] \E[a',b'] = \E[a,b]$ whenever $[a',b'] \subseteq [a,b]$;
    \item $\mathrm{Ran} \E[1,a_{j+1}] = \mathrm{Ran} \E[1, a_{j}] \cap \mathrm{Ran} \E[a_{j-1}+1,a_{j+1}]$ for each $j \in 1 \dots J$, where $\mathrm{Ran}$ denotes the range or output subalgebra of a conditional expectation.
    \item For each $j = 2 \dots J$, Assumption \ref{assump:glue} for $\E[a_{j-1}+1, a_{j+1}]$ and $\E[1, a_{j}]$ with gluing coefficient $\alpha \leq g(\E[a_{j-1}+1, a_{j+1}], \E[1, a_{j}])$, and $g(\E[a_{j-1}+1, a_{j+1}], \E[1, a_{j}]) \geq 1$.
\end{enumerate}
\begin{lemma}[Entropy Decay Recomposition] \label{lem:chainmin-strong}
    Assume (1)-(4). Then
    \begin{equation}
    \begin{split}
        & \sum_{j=1}^{J}  D(\Phi[1, a_{j-1}] (\rho)  \| \E[a_{j-1}+1, a_{j+1}](''))
         \\ & \geq \bigg ( \prod_{j=2}^J g(\E[a_{j-1}+1, a_{j+1}], \E[1, a_{j}]) \bigg )^{-1}
            D(\Phi[1, m] (\rho) \| \E[1,m] ('') )
        \pl.
    \end{split}
    \end{equation}
\end{lemma}
Lemma \ref{lem:chainmin-strong} is proven as Lemma \ref{lem:chainmin-strong-supp} in the Supplemental Material. This Section's main Theorem combines two `streams' of iterated relative entropies from the same sequence of channels:
\begin{theorem} \label{thm:compose-2stream}
    Assume (1)-(4) and that $\Phi[a_{j-1} + 1, a_{j+1}]$ has $\lambda$-SDPI toward $\E[a_{j-1} + 1, a_{j+1}]$ for each $j = 1 \dots J$. Then $\Phi[1,m]$ has $\lambda'$-SDPI with
    \begin{equation}
        \lambda' \geq 1 - \bigg ( 1 +
        \frac{\lambda }{2} \bigg ( \prod_{j=2}^J g(\E[a_{j-1}+1, a_{j+1}], \E[1, a_{j}]) \bigg )^{-1} \bigg )^{-1} \pl.
    \end{equation}
    If the local SDPI constant $\lambda$ and the gluing constants hold under arbitrary tensor extensions, then $\Phi[1,m]$ has $\lambda'$-CSDPI.
\end{theorem}
\begin{proof}
    When $b < a$, let $\Phi[a,b] := \Id$. Let $a_{J+2} = a_{J+1}$, and $a_{-1} = a_0$. Consider the two rewritings
    \begin{equation}
    \begin{split}
        & \Phi[1,m] =
        \Phi[a_{2 \lceil J/2 \rceil} + 1, m] \circ \left ( \overset{\leftarrow}{\prod_{j=0}^{\lceil J/2 \rceil - 1}}
         \Phi[a_{2 j} + 1, a_{2 (j + 1)}] \right )
        \\ & = 
        \Phi[a_{2 \lfloor J / 2 \rfloor +  1} + 1, m] \circ 
        \left ( \overset{\leftarrow}{\prod_{j=0}^{\lfloor J/2 \rfloor - 1}}
        \Phi[a_{2 j + 1} + 1, a_{2 j + 3}] \right ) \circ \Phi[1, a_1]
        \pl.
    \end{split}
    \end{equation}
    Via Lemma \ref{lem:chainmin1}, using the data processing inequality to ignore the final channels when $J$ is even,
    \begin{equation} \label{eq:chain-split-1}
    \begin{split}
        & D(\rho \| \E[1,m](\rho)) - D(\Phi[1,m] (\rho) \| \E[1,m](\rho)) 
        \\ \geq & \lambda \sum_{j=0}^{\lceil J/2 \rceil - 1}  D \bigg (
            \overset{\leftarrow}{\prod_{s=0}^{j-1}}
            \Phi[a_{2 s} + 1, a_{2 (s + 1)}] (\rho)
         \bigg \| \E[a_{2 j} + 1, a_{2 (j + 1)}]('') \bigg )
        \\ =  & \lambda \sum_{j=0}^{\lceil J/2 \rceil - 1}  D \big (
            \Phi[1, a_{2 j}](\rho)
        \big \| \E[a_{2 j} + 1, a_{2 (j + 1)}]('') \big ) \pl.
    \end{split}
    \end{equation}
    Similarly, ignoring the first term and ignoring the last channels when $J$ is odd.
    \begin{equation} \label{eq:chain-split-2}
    \begin{split}
        & D(\rho \| \E[1,m](\rho)) - D(\Phi[1,m] (\rho) \| \E[1,m](\rho)) 
        \\ \geq & \lambda \sum_{j=0}^{\lfloor J/2 \rfloor - 1}  D \bigg (
            \bigg ( \overset{\leftarrow}{\prod_{s=0}^{j-1}}
            \Phi[a_{2 s + 1} + 1, a_{2 s + 3}] \circ \Phi[1, a_1] \bigg ) (\rho)
        \bigg \| \E[a_{2 j + 1} + 1, a_{2 j + 3}]('') \bigg ) 
        \\ = & \lambda \sum_{j=0}^{\lfloor J/2 \rfloor - 1}  D \big (
            \Phi[1, a_{2 j + 1}](\rho)
        \big \| \E[a_{2 j + 1} + 1, a_{2 j + 3}]('') \big ) \pl.
    \end{split}
    \end{equation}
    Adding the equations above,
    \begin{equation} \label{eq:2streams}
    \begin{split}
        & D(\rho \| \E[1,m](\rho)) - D(\Phi[1,m] (\rho) \| \E[1,m](\rho))  \geq
        \\ & \frac{\lambda}{2} \sum_{j=1}^{J}
            D ( \Phi[1, a_{j-1}](\rho) \| \E[a_{j-1} + 1, a_{j+1}]('')) \pl.
    \end{split}
    \end{equation}
    Applying Lemma \ref{lem:chainmin-strong} yields that
    \begin{equation}
    \begin{split}
        & D(\rho \| \E[1,m](\rho))
        \geq \bigg ( 1 +
        \frac{\lambda }{2} \bigg ( \prod_{j=2}^J g(\E[a_{j-1}+1, a_{j+1}], \E[1, a_{j}]) \bigg )^{-1} \bigg )
        D(\Phi[1, m] (\rho) \| \E[1,m] ('') )  \pl,
    \end{split}
    \end{equation}
    finishing the SDPI bound after re-arranging terms. One obtains CSDPI by extending each channel and associated multiplicative domain projection by the identity on an auxiliary system of arbitrary dimension.
\end{proof}

\section{Conclusions and Outlook}
This work addresses an open question noted in \cite{schuster_random_2025} and \cite{laracuente_approximate_2024}: whether shallow $k$-design convergence applies to `unstructured' or time-homogeneous random circuits such as 1-D brickwork. Furthermore, even compared to the structured schemes of \cite{laracuente_approximate_2024, schuster_random_2025}, the $1/\exp(O(\log^*n))$ entropy convergence rate is faster than the $1/O(\log n)$ that would result from straightforwardly applying multiplicative error to entropy conversion to bounds therein. A lingering open question is whether unstructured random circuits yield shallow multiplicative-error designs.

\section{Acknowledgments}
I thank Felix Leditzky for helpful feedback on an early version of some results and Nick Hunter-Jones for helpful conversations during a visit to the University of Texas at Austin. I thank Shivan Mittal for pointing out an erroneous proof step in a previous version. I thank Angela Capel for noting reference \cite{ohya_quantum_1993}.

ChatGPT Pro versions 5.5 and 5.6 Sol were used to inspire Lemmas \ref{lem:log-iterate-appended} and \ref{lem:log-compare-appended}, which appear in the Supplementary Information. These tools were used to suggest simplified bounds on the quantities of interest and ideas for proof, but the proofs are the work of the author. The AI tools were also used to suggest corrections in editing and revision. No facts, equations, or derivations were copied from AI tools without independent verification by the author. The author takes full responsibility for the manuscript's content.

\section{Data Availability Statement}
No new data were created or analyzed.

\printbibliography

\appendix


\section{Supplemental Material: Technical Background, Lemmas, and Proofs}
\subsection{Mathematical Estimates}
\begin{lemma} \label{lem:log-iterate-appended}
    Assume $x > 1$, and $R_j(y) \leq a \ln y + b$ for fixed $a > 0$ and $b \geq 1$, each $j = 1, \dots, \log^*x$, and every $y \geq 1$. Assume each $R_j$ is non-decreasing in its input. Then
        $R_{\log^*(x)} \circ \dots \circ R_1(x) \leq 2 \tilde{a} + (1+\tilde{a}) (\ln \tilde{a} +
            \max \{ b /\tilde{a}, \tilde{a}^2 \})$, where $\tilde{a} = \max\{1,a\}$.
\end{lemma}
\begin{proof}
    The key is the recursion relation for $j$ such that $\ln^j x \geq 1$,
    \begin{equation} \label{eq:iter-log-recur-1}
    \begin{split}
        R(a' \ln^j x + b') & \leq a \ln (a' \ln^j x + b') + b
        \\ & = a ( \ln a' + \ln^{(j+1)} x + \ln (1 + b' / a' \ln^j x) ) + b
        \\ & \leq a ( \ln a' + \ln^{(j+1)} x + \ln (1 + b'/a') ) + b
        \\ & = a \ln^{(j+1)} x + a (\ln a' + \ln(1+b'/a')) + b
    \end{split}
    \end{equation}
    Note that the upper-bound on $R(x)$ is monotonic in $x$, so it suffices to overestimate each round. At each step of the recursion, we may substitute $a' \mapsto a$, $b' \mapsto a (\ln a' + \ln(1+b'/a')) + b$, and $x \mapsto \ln x$ for such an overestimate. When $b' = a^3$,
    \begin{equation}
        \frac{a + b' / a}{1+a} = \frac{a(1+a)}{1+a} = a \geq  \ln (1 + a^2) = \ln(1 + b'/a) \pl.
    \end{equation}
    Note the derivative in $b'$ of the difference,
    \begin{equation}
        \frac{1}{a(1+a)} - \frac{1}{a(1+b'/a)} \pl,
    \end{equation}
    is non-negative when $b' \geq a^3$. Henceforth, assume $b \geq a^3$. Therefore, if in a given step of the recursion we map $b' \rightarrow b''$, then
    \begin{equation}
    \begin{split}
        b'' & \leq a \Big ( \ln a + \frac{a + b'/a}{1+a} \Big ) + b
         = \frac{1}{1+a} b' + \frac{a^2}{1+a} + a \ln a + b \pl.
    \end{split}
    \end{equation}
    Let $r := 1/(1+a)$, and $c = a^2/(1+a) + a \ln a + b$, so the recursion reduces to $b' \rightarrow r b' + c$. Iterating yields that after $n$ rounds of recursion,
    \begin{equation}
        b' \mapsto r^n b' + \sum_{j=0}^{n-1} r^j c \leq r^n b' + \frac{c(1-r^n)}{1-r} = \Big ( \frac{1}{1+a} \Big )^n  \Big ( b' - (1+a) c / a \Big ) + (1+a) c / a \pl.
    \end{equation}
    via the geometric series formula. The first term is always negative, so the final value is at most $(1+a) c / a$. The derivation holds as long as $n \leq \log^* x - 1$, so that after the final step, $x \mapsto a \ln^{n+1}x + y_n$ such that $\ln^{n+1} x \leq 1$ when $n = \log^* x - 1$, and $y_n \leq (1+a) c / a$. This yields an upper bound of
    \begin{equation}
    \begin{split}
        & a + \frac{1+a}{a} \big ( a^2/(1+a) + a \ln a + b \big ) 
            = 2 a + (1+a) \ln a + (1+a) b /a \pl.
    \end{split}
    \end{equation}
    If $a < 1$, then observing that $y \mapsto a \ln y + b$ is non-decreasing in $a$ for $y \geq 1$, one may substitute 1. If the initial input and $b$ are both at least 1, then every intermediate value of the upper-bound is at least 1. If $b < a^3$, then we observe that the end value is increasing in $b$, so one may substitute $b = a^3$. 
\end{proof}
\begin{lemma} \label{lem:iterate-inverse}
    Let $R_1, \dots, R_m$ be functions on the real numbers. 
    Let $W_1, \dots, W_m$ be non-decreasing functions on the real numbers such that for each $j = 1 \dots m$, $W_j(R_j(x)) \geq x$ for every $x$. If $R_m \circ \dots \circ R_1(x) \leq c$, then $W_1 \circ \dots \circ W_m(c)\geq x$.
\end{lemma}
\begin{proof}
    For each $j = 1 \dots m$, if $c_j \geq R_j \circ \dots \circ R_1(x)$, then  $W_j(c_j) \geq R_{j-1} \circ \dots \circ R_1(x)$. Starting from $c_{m} = c \geq R_m \circ \dots \circ R_1(x)$, define $c_{j-1} = W_j(c_j)$ for each $j = m \dots 1$ and apply this implication inductively.
\end{proof}
\begin{lemma} \label{lem:log-compare-appended}
    Assume $x, a, b > 0$. If $\ln a + b \leq 1$ or if $\ln a + b > 1$ and $x \geq a (\ln a + b + \ln (2 (\ln a + b)))$, then $x \geq a (\ln (x) + b)$.
\end{lemma}
\begin{proof}
    The function $f(x) := x - a (\ln (x) + b)$ is minimized at $x = a$, where $f(a) = a (1 - \ln a - b)$, so its derivative thereafter is positive. If $\ln a + b \leq 1$, then $f(a) \geq a - a = 0$. Therefore, attention is needed only when $\ln a + b > 1$. At the given threshold,
    \begin{equation}
    \begin{split}
        & f(a (\ln a + b + \ln (2 (\ln a + b))))
        \\ & = a (\ln a + b + \ln (2 (\ln a + b))) - a \ln (a (\ln a + b + \ln (2 (\ln a + b)))) - a b
        \\ & = a \ln (2 (\ln a + b)) - a \ln(\ln a + b + \ln (2 (\ln a + b))) \pl.
    \end{split}
    \end{equation}
    The condition then reduces to that $\ln (2 (\ln a + b)) = \ln 2 + \ln(\ln a + b) \leq \ln a + b$. Since $\ln a + b > 1$, $\ln(\ln a + b) \leq \ln a + b - 1$. Therefore, $\ln 2 + \ln (\ln a + b) \leq \ln a + b + (\ln 2 - 1 ) \leq \ln a + b$.
\end{proof}
\subsection{Base Estimates for Entropic Convergence of Small Random Circuits}
For a channel $\Phi$, let $\Phi^*$ denote its adjoint with respect to the trace, defined by the relation that $\tr(\Phi(X) Y) = \tr(X \Phi^*(Y))$ for every pair of input operators $X,Y$. Resulting from \cite[Theorem 2.5]{gao_complete_2025}, if $\Phi$ is a unital quantum channel, $\E$ its multiplicative domain projection, and $t \in \NN$ such that $0.9 \E \prec (\Phi^* \Phi)^t \prec 1.1 \E$, then
\begin{equation} \label{eq:iterate-rel}
    D( \Phi\otimes \Id (\rho)||(\Phi\circ \E)\otimes \Id (\rho))\le \Big (1-\frac{1}{2 t} \Big ) D(\rho|| \E \otimes \Id (\rho)) \pl,
\end{equation}
where $\Id$ is the identity on an auxiliary system of arbitrary, finite dimension.
\begin{lemma} \label{lem:brickworkknown}
     Let $\Phi_{BR}$ denote the channel given by applying a 1-D brickwork random circuit with locally Haar random unitaries. Then
     \begin{equation}
     \begin{split}
         & (1-\epsilon) \Phi_{\mathrm{Haar}, k} \prec \Phi_{BR}^{\lceil (2 n k + \log_2 (1/\epsilon)) / \log_2(1/(1-2^{-53})) \rceil} \prec (1+\epsilon) \Phi_{\mathrm{Haar}, k} \pl, \text{ and }
         \\ & (1-\epsilon) \Phi_{\mathrm{Haar}, k} \prec (\Phi_{BR}^* \Phi_{BR})^{\lceil (2 n k + \log_2 (1/\epsilon)) / (2 \log_2(1/(1-2^{-53}))) \rceil} \prec (1+\epsilon) \Phi_{\mathrm{Haar}, k} \pl.
     \end{split}
     \end{equation}
\end{lemma}
\begin{proof}
    The recent result of \cite{baer_random_2026} estimates that the spectral gap for 1-D brickwork is at least $2^{-53}$. Via \cite[Lemma 4]{brandao_local_2016}, $\Phi_{BR}^{m}$ forms an $\epsilon$-approximate relative error $k$-design when $\epsilon \geq 2^{2 n k} (1-2^{-53})^m$.

    The adjoint of $\Phi_{BR}$ with respect to the trace, denoted $\Phi_{BR}^*$, is just a brickwork random circuit with the layers reversed. Therefore, it has the same spectral gap, yielding the second comparison.
\end{proof}
Recall \cite{belkin_approximate_2024}. Based on results shown therein, we prove the following Lemma:
\begin{lemma} \label{lem:parallel-linear}
    Every $n$-qudit, 2-local, $\ell$-layer, connected parallel random circuit architecture with local dimension $q$ that applies Haar-random local gates induces a random unitary measure $\mu$ that is a $\lambda$-converging entropic design with
    \begin{equation}
        \lambda \geq 1 / (2\lceil (\ln q)(2 k n + \log_q (10)) (4 C(\sqrt{q},k))^{\ell - 1} \rceil) \pl.
    \end{equation}
    Also, $\Phi_{\mu, k}^m \precsucc{\epsilon}{} \Phi_{\mathrm{Haar}, k}$ in multiplicative error whenever
    \begin{equation} \label{eq:rel-err-base}
        m \geq \lceil 2 (\ln q)(2 k n + \log_q (1/\epsilon)) (4 C(\sqrt{q},k))^{\ell - 1} \rceil \pl.
    \end{equation}
    A general formula for $C(\sqrt{q},k)$ is given in \cite{belkin_approximate_2024}. In particular, $C(\sqrt{4},k) = C(2,k) = O(1)$ in $k$. More broadly, $C(\sqrt{q},k) = O(\mathrm{poly}(k))$.
\end{lemma}
\begin{proof}
    By the assumptions on $\ell$-layer parallel random circuits, $\Phi$ has the Haar-weighted $k$-fold twirl as its multiplicative domain projection (as confirmed by the non-trivial convergence estimates in \cite{belkin_approximate_2024}). In particular, the fixed point algebra of the full circuit is the intersection of local fixed point algebras. When the circuit is connected, universality of 2-qudit unitaries implies that this fixed point must be the same as that of the global unitary group.
    
    Recall \cite[Equations (51) and (55)]{belkin_approximate_2024}, stating that the spectral gap $1 - s_*$ of an arbitrary $\ell$-layer parallel random circuit is bounded as
    \begin{equation}
        s_*^2 \leq 1 - (1 - e^{-1/2 C(\sqrt{q}, k)})^{\ell - 1} \pl,
    \end{equation}
    where $C(\sqrt{q}, k)$ depends on the convergence of 1-D brickwork and is calculated for several cases therein. When $C(\sqrt{q}, k)$ is large enough (and we may enlarge it by a constant to ensure so), we may estimate
    \begin{equation}
        - \ln s_* \geq \frac{1}{2 (4 C(\sqrt{q}, k))^{\ell - 1}} \pl.
    \end{equation}
    If the original random circuit is $\Phi$, then the same spectral gap estimate applies to $\Phi^*$, its adjoint under the trace, since the adjoint reverses gate order while preserving the essential properties of an $\ell$-layer parallel random circuit. Hence, $\Phi^* \Phi$ has spectral gap $1-g$ with $g = s_*^2$. By \cite[Lemma 4]{brandao_local_2016}, $(\Phi^* \Phi)^m \precsucc{\epsilon}{} \Phi_{\mathrm{Haar}, k}$ when $\epsilon \geq q^{2 k n} g^{m}$ after $m$ applications. Therefore, for a given $\epsilon > 0$, it suffices to take
    \begin{equation}
        m \geq \log_{g} (\epsilon q^{- 2 k n}) = \log_{s_*}(\epsilon q^{- 2 k n}) / 2 \pl.
    \end{equation}
    Dividing logarithms to change the base to $q$ yields the sufficient condition that
    \begin{equation}
        m \geq (\ln q) (2 k n + \log_q (1/\epsilon)) (4 C(\sqrt{q}, k))^{\ell - 1} \pl.
    \end{equation}
    Equation \eqref{eq:rel-err-base} follows from similar reasoning, but using $\Phi$ itself instead of $\Phi^* \Phi$.
    
    The value of $C(\sqrt{q}, k)$ given for general architectures comes from \cite{belkin_approximate_2024}. Specifically, $C(q,k)$ is defined such that $C(q,k)(2 n k + \ln(1/\epsilon))$ bounds the convergence rate of an $\epsilon$-approximate $k$-design for 1-D brickwork with $q$-dimensional qudits. The value for $q=4$ comes from applying the same procedure to the asymptotic bound from \cite{chen_incompressibility_2024} or the recent \cite{baer_random_2026}. A general $O(\mathrm{poly}(k))$ (ignoring dependence on $q$ or $\ell$) bound for $C(r,k)$ when $r^2 \geq 2$ is an integer (but not necessarily $r$) is noted in \cite[Section I.F]{belkin_approximate_2024}, bounding $C(\sqrt{q}, k)$ for any positive integer $q \geq 2$. 
\end{proof}
In principle, recent spectral gap estimates for 1-D brickwork \cite{baer_random_2026} should yield a concrete bound on $C(\sqrt{q},k)$ that is $O(1)$ in $k$. This does yield a bound for $q=4$. However, extending this bound to ``incomplete'' parallel $\ell$-layer circuits for $q = 2$ requires tensor network computations as detailed in \cite{belkin_approximate_2024}. 
\begin{lemma} \label{lem:small-random}
    Let $G$ be a simple, connected, undirected graph, of degree at most $c$, and of $n$ nodes. Assuming each edge applies a channel with at least $\lambda'$-CSDPI toward a pair-local Haar twirl, the channel $\Phi_{G, k}$ has CSDPI constant
    \begin{equation}
        \lambda \geq \beta_\epsilon \lambda' /  ( n c \lceil 2 (\ln q) (4 C(\sqrt{q}, k))^{c} (2 k n + \log_q (1/\epsilon)) \rceil) \pl,
    \end{equation}
    for given $\epsilon$ for which $0 < \epsilon < 1/12$, where $C(\sqrt{q}, k)$ is the constant as in Lemma \ref{lem:parallel-linear}, and $\beta_\epsilon$ is defined as in Equation \eqref{eq:quasifac}. In the particular case that $G$ is the cycle graph or path graph with $q=2$,
    \begin{equation}
        \lambda_G \geq \beta_\epsilon \lambda' /  ( n c \lceil (2 n k + \log_2(1/\epsilon)) / \log_2(1/(1-2^{-53})) \rceil) \pl.
    \end{equation}
\end{lemma}
\begin{proof}
    By Vizing's Theorem, the graph admits an edge coloring using $\ell \leq c+1$ colors. Let $\nu$ be the unitary measure induced by the 2-local, $\ell$-layer parallel random circuit that for each colored edge applies a locally Haar-random gate to the nodes defined by the edge at the layer defined by that color. For each edge $(a,b) \in G$, let $\Phi_{a,b,k}$ denote the random $k$-fold gate application or twirl to the associated pair of qudits, and $\E_{a,b,k}$ its individual conditional expectation or Haar twirl. For a channel $\Psi$, denote by $\tilde{\Psi} = \Id \otimes \Psi$ its extension to an auxiliary system of arbitrarily large dimension. Iterating Equation \eqref{eq:chain-iter} for any $m \in \NN$,
    \begin{equation}
        m \sum_{(a,b) \in G} D(\rho \| \tilde{\E}_{a,b,k}(''))
            \geq D (\rho \| \tilde{\Phi}_{\nu, k}^m (\rho) ) \pl,
    \end{equation}
    Let $m_0 := \lceil 2 (\ln q) (2 k n + \log_q (1/\epsilon)) (4 C(\sqrt{q}, k))^{\ell - 1} \rceil$with $C(\sqrt{q}, k)$ defined as therein. By Lemma \ref{lem:parallel-linear}, $\Phi^{m_0}_{\nu, k}$ is the channel given by an $\epsilon$-approximate multiplicative error $k$-design, and
    \begin{equation}
        m_0 \sum_{(a,b) \in G} D(\rho \| \tilde{\E}_{a,b,k}(''))
            \geq \beta_{\epsilon} D (\rho \| \tilde{\Phi}_{\mathrm{Haar}, k}(\rho) ) \pl.
    \end{equation}
    By Lemma \ref{lem:decay-merge}, noting that $|E(G)| \leq c n / 2$,
    \begin{equation} \label{eq:lambdaG-inproof}
        \lambda_G \geq \frac{\beta_\epsilon \lambda'}{c n m_0}
            \geq \beta_\epsilon \lambda' / n c \lceil 2 (\ln q) ( 4 C(\sqrt{q}, k))^{c} (2 k n + \log_q (1/\epsilon)) \rceil \pl.
    \end{equation}
    For a path graph, which is necessarily 2-colorable, $\prod_{(a,b) \in E(G)} \tilde{\E}_{a,b,k}$ can be ordered to produce 1-D brickwork with Haar local gates and open boundary conditions. By Lemma \ref{lem:brickworkknown}, an $\epsilon$-approximate relative $k$-design is achieved with
    \begin{equation}
        m_{\mathrm{path}} = \lceil (2 n k + \log_2(1/\epsilon)) / \log_2(1/(1-2^{-53})) \rceil \pl.
    \end{equation}
    This $m_{\mathrm{path}}$ replaces $m_0$ in Equation \eqref{eq:lambdaG-inproof}. A cycle graph is an average over each path graph given by removing one edge, so as relative-error convergence is convex, the same bounds apply.
\end{proof}
\subsection{Proofs of Main Lemmas and Theorems}
Recall a form of inequality used to combine entropy decay estimates \cite[Corollary II.15]{laracuente_quasi-factorization_2022} \cite[Lemma 2.3]{gao_complete_2025}:
for quantum channels $\E$ and $\Psi$ such that $\Psi \E = \E$, density $\text{supp}(\rho) \subseteq \text{supp}(\E(\rho))$, $\epsilon \in (0,1)$, if
$(1 - \epsilon) \E \prec \Psi \prec (1 + \epsilon) \E$,
then with $\beta_{\epsilon} \geq 1 - 12 \epsilon$,
\begin{equation} \label{eq:quasifac}
    D(\rho \| \Psi(\rho)) \geq \beta_{\epsilon} D(\rho \| \E(\rho)) \pl.
\end{equation}
\begin{lemma} \label{lem:decay-merge-supp}
    Let $\Phi = \sum_{i=1}^\ell p_i \Phi_i$ be a quantum channel with fixed point projection $\E$ for which each $\Phi_i$ has $\lambda_i$-CSDPI toward multiplicative domain projection $\E_i$, $\E \E_i = \E_i \E = \E$, $\E_i \Phi_i = \Phi_i \E_i = \E_i$, and \ref{assump:quasifac} holds with constants $(\alpha_i)_{i=1}^\ell$. Then $\Phi$ has $\min_i p_i \lambda_i / \alpha_i$-CSDPI.
\end{lemma}
\begin{proof}
    Via the convexity of quantum relative entropy, Equation \eqref{eq:chain-rule}, and local CSDPI assumption, 
    \begin{equation}
    \begin{split}
        & D((\Id \otimes \Phi)(\rho) \| (\Id \otimes \E)(''))
        \\ & = D \left ( \sum_{i} p_i (\Id \otimes \Phi_i)(\rho) \bigg \| (\Id \otimes \E)('') \right )
        \\ & \leq \sum_{i} p_i D \left ( (\Id \otimes \Phi_i)(\rho)
            \| (\Id \otimes \E)('') \right )
        \\ & = \sum_{i} p_i  \Big (D \left ( (\Id \otimes \Phi_i)(\rho) \| (\Id \otimes \E_i)('') \right )
            + D((\Id \otimes (\E_i \circ \Phi_i))(\rho) \| (\Id \otimes \E)('')) \Big )
        \\ & \leq \sum_{i} p_i \Big (  (1 - \lambda_i)  D \left ( \rho
            \| (\Id \otimes \E_i)('') \right ) + D((\Id \otimes \E_i)(\rho) \| (\Id \otimes \E)('')) \Big )
        \\ & = D(\rho \| \E(\rho)) - \sum_i \lambda_i p_i D \left ( \rho \| (\Id \otimes \E_i)('') \right ) \pl.
    \end{split}
    \end{equation}
    Applying Assumption \ref{assump:quasifac},
    \begin{equation}
        \sum_i \lambda_i p_i D \left ( \rho \| (\Id \otimes \E_i)('') \right )
            \geq (\min_i \lambda_i p_i / \alpha_i) D(\rho \| \E(\rho)) \pl,
    \end{equation}
    completing the Lemma.
\end{proof}
\begin{lemma} \label{lem:chainmin1-supp}
Let $(\Phi_j)_{j=1}^m$ be a given family of quantum channels with respective $(\lambda_j)_{j=1}^m$-SDPI to multiplicative domain projections $(\E_j)_{j=1}^m$ such that $\Phi_j \E_j = \E_j \Phi_j$ for each $j$. Let $\E$ be a trace-preserving conditional expectation commuting with each $\Phi_j$ for which $\E \E_j = \E_j \E = \E$. Let $(\Psi_j)_{j=1}^m$ be a family of channels each commuting with $\E$. Then
\begin{equation*}
    \begin{split}
         D(\rho \| \E(\rho)) - D((\Psi_m \Phi_m) \dots (\Psi_1 \Phi_1) (\rho) \| \E (''))
            \mbnline \mbalign \geq \sum_{j=1}^m \lambda_j D( (\Psi_{j-1} \Phi_{j-1}) \dots (\Psi_1 \Phi_1)(\rho) \| \E_j ('')) \pl,
    \end{split}
\end{equation*}
where $(\Psi_{0} \Phi_0) \dots (\Psi_1 \Phi_1) := \Id$.
\end{lemma}
\begin{proof}
    First, using the data processing inequality,
    \begin{equation} \label{eq:chain1}
        D((\Psi_m \Phi_m) \dots (\Psi_1 \Phi_1) (\rho) \| \E ('')) 
            \leq D(\Phi_m (\Psi_{m-1} \Phi_{m-1}) ... (\Psi_1 \Phi_1) (\rho) \| \E ('')) \pl,
    \end{equation}
    Using the chain rule of relative entropy, Equation \eqref{eq:chain-rule},
    \begin{equation} \label{eq:chain2}
    \begin{split}
        \mbalign D(\Phi_m (\Psi_{m-1} \Phi_{m-1}) ... (\Psi_1 \Phi_1) (\rho) \| \E (''))
            \mbnline & =  D(\Phi_m (\Psi_{m-1} \Phi_{m-1}) ... (\Psi_1 \Phi_1) (\rho) \| \E_m ('')))
            \\ &  + D(\E_m \Phi_m (\Psi_{m-1} \Phi_{m-1}) ... (\Psi_1 \Phi_1) (\rho) \| \E (''))) \pl, 
    \end{split}
    \end{equation}    
    Then using the assumed $\lambda_m$-SDPI,
    \begin{equation} \label{eq:chain3}
        \mbalign D(\Phi_m (\Psi_{m-1} \Phi_{m-1}) ... (\Psi_1 \Phi_1)(\rho) \| \E_m ('')))
        \mbnline \mbalign \leq
            (1 - \lambda_m) D( (\Psi_{m-1} \Phi_{m-1}) \dots (\Psi_1 \Phi_1)(\rho) \| \E_m('')) \pl .
    \end{equation}
    By the definition of the multiplicative domain there exists a map $\Phi_m^R$ such that $\Phi_m^R \circ\Phi_m\circ\E_m = \E_m$, and in addition we also have $\Phi_m^R \circ\Phi_m\circ\E = \E$ since $\E = \E_m \E = \E \E_m$. Then, since $\E_m\circ\Phi_m = \Phi_m\circ\E_m$ and $\E \circ\Phi_m = \Phi_m\circ\E$, we have by data-processing in both directions that
    \begin{equation} \label{eq:chain4}
        D(\E_m \Phi_m (\Psi_{m-1} \Phi_{m-1}) ... (\Psi_1 \Phi_1) (\rho) \| \E ('')))
            = D(\E_m (\Psi_{m-1} \Phi_{m-1}) ... (\Psi_1 \Phi_1) (\rho) \| \E (''))) \pl.
    \end{equation}
    Recombining Equations \eqref{eq:chain1}, \eqref{eq:chain2}, \eqref{eq:chain3}, and \eqref{eq:chain4},
    \begin{equation}
    \begin{split}
        & D((\Psi_m \Phi_m) \dots (\Psi_1 \Phi_1) (\rho) \| \E (''))
            \\ & \leq (1 - \lambda_m) D( (\Psi_{m-1} \Phi_{m-1}) \dots (\Psi_1 \Phi_1)(\rho) \| \E_m('')) + D(\E_m (\Psi_{m-1} \Phi_{m-1}) \dots (\Psi_1 \Phi_1) (\rho) \| \E (''))) \pl.
    \end{split}
    \end{equation}
    Using Equation \eqref{eq:chain-rule} to recombine the terms on the right-hand side,
    { \begin{equation} \label{eq:chain5}
    \begin{split}
         D((\Psi_m \Phi_m) \dots (\Psi_1 \Phi_1) (\rho) \| \E (''))
        \mbnline \mbalign & \leq D((\Psi_{m-1} \Phi_{m-1}) \dots (\Psi_1 \Phi_1) (\rho) \| \E (''))
        \\ & - \lambda_m D( (\Psi_{m-1} \Phi_{m-1}) \dots (\Psi_1 \Phi_1)(\rho) \| \E_m ('')).
    \end{split}
    \end{equation} }
    Iterating \eqref{eq:chain5},
    { \begin{align}
        & D(\rho \| \E(\rho)) - D((\Psi_m \Phi_m) \dots (\Psi_1 \Phi_1) (\rho) \| \E (''))
            \mbnline \mbalign \geq \sum_{j=1}^m \lambda_j D( (\Psi_{j-1} \Phi_{j-1}) \dots (\Psi_1 \Phi_1)(\rho) \| \E_j ('')) .
    \end{align} }
    This completes the proof.
\end{proof}
\begin{lemma} \label{lem:chainmin-strong-supp}
    Consider a sequence of channels $\Phi[1], \dots, \Phi[m]$, where $\Phi[a,b]$ denotes $\Phi[b] \circ \dots \circ \Phi[a]$, and $\E[a,b]$ denotes the multiplicative domain projection of $\Phi[a,b]$. Let $\Phi[a,b] = \Id$ when $b < a$, and $\Phi[a] = \Id$ whenever $a < 1$ or $a > m$.  Let $J \in \NN$ and $a_0, \dots, a_{J+1}$ be a strictly increasing sequence for which $a_0 = 0$, and $a_{J+1} = m+1$. Assume that...
    \begin{itemize}
        \item $\Phi[x] \E[a,b] = \E[a,b] \Phi[x] = \E[a,b]$ whenever $x \in [a,b]$.
        \item $\E[a',b'] \E[a,b] = \E[a,b] \E[a',b'] = \E[a,b]$ whenever $[a',b'] \subseteq [a,b]$.
        \item For each $j \in 1 \dots J$, $\mathrm{Ran} \E[1,a_{j+1}] = \mathrm{Ran} \E[1, a_{j}] \cap \mathrm{Ran} \E[a_{j-1}+1,a_{j+1}]$, where $\mathrm{Ran}$ denotes the range or output subalgebra of a conditional expectation.
        \item For each $j = 2 \dots J$, \ref{assump:glue} for $\E[a_{j-1}+1, a_{j+1}]$ and $\E[1, a_{j}]$ with gluing coefficient $\alpha \leq g(\E[a_{j-1}+1, a_{j+1}], \E[1, a_{j}])$, and $g(\E[a_{j-1}+1, a_{j+1}], \E[1, a_{j}]) \geq 1$.
    \end{itemize}
     Then
    \begin{equation} \label{eq:chainmin-strong}
    \begin{split}
        & \sum_{j=1}^{J}  D(\Phi[1, a_{j-1}] (\rho)  \| \E[a_{j-1}+1, a_{j+1}](''))
         \\ & \geq \bigg ( \prod_{j=2}^J g(\E[a_{j-1}+1, a_{j+1}], \E[1, a_{j}]) \bigg )^{-1}
            D(\Phi[1, m] (\rho) \| \E[1,m] ('') )
        \pl.
    \end{split}
    \end{equation}
\end{lemma}
\begin{proof}
    The proof is inductive. At each $j$th step for $j = 1 \dots J$, we show that $A_j + B_j \geq g_j^{-1} A_{j+1}$ for some scalar $g_j$, where
    \begin{equation} \label{eq:induct-sequence}
        A_j = D(\Phi[1, a_{j-1}](\rho) \| \E[1, a_{j}]('')) \pl,
    \end{equation}
    and
    \begin{equation}
        B_j = D(\Phi[1,a_{j-1}](\rho) \| \E[a_{j-1}+1, a_{j+1}]('')) \pl.
    \end{equation}
    Observe that the sum on the left-hand side of Equation \eqref{eq:chainmin-strong} equals $\sum_{j=1}^j B_j$. As the base case, $A_2 = D(\Phi[1, a_{1}](\rho) \| \E[1, a_{2}]('')) \leq B_1$. Hence, one may assume $g_1 = 1$. Using the data processing inequality, for each $j = 2 \dots J$,
    \begin{equation} \label{eq:chain-range-0}
    \begin{split}
        & A_j
        \geq D(\E[a_{j-1} + 1, a_j] \Phi[1, a_{j-1}](\rho) \| \E[1, a_j](''))
        \pl.
    \end{split}
    \end{equation}
    Using the data processing inequality and commutation with decoherence-free subspace projections to apply $\Phi[a_{j-1}+1, a_{j}]$ followed by $\E[a_{j-1}+1, a_{j+1}]$ to both arguments,
    \begin{equation} \label{eq:chain-range-1}
    \begin{split}
        & D(\E[a_{j-1} + 1, a_j] \Phi[1, a_{j-1}](\rho) \| \E[1, a_j](''))
        \\ & = D(\Phi[a_{j-1}+1, a_{j}] \E[a_{j-1} + 1, a_j] \Phi[1, a_{j-1}](\rho) \| 
                \Phi[a_{j-1}+1, a_{j}] \E[1, a_{j}] \Phi[1, a_{j-1}] (\rho) )
        \\ & = D(\E[a_{j-1} + 1, a_j] \Phi[1, a_{j}](\rho) \| \E[1, a_{j}](''))
        \\ & \geq D(\E[a_{j-1}+1, a_{j+1}] \Phi[1, a_{j}] (\rho) \| 
            \E[a_{j-1}+1, a_{j+1}] \E[1, a_{j}] \Phi[1, a_{j}](\rho) ) 
        \pl.
    \end{split}
    \end{equation}
    Using the chain rule and data processing inequality
    to apply $\Phi[a_{j-1}+1, a_j]$ to both arguments of $B_j$,
    \begin{equation} \label{eq:chain-range-2}
    \begin{split}
        B_j & = D(\Phi[1,a_{j-1}](\rho) \| \E[a_{j-1}+1, a_{j}](''))
            \\ & \pl \pl \pl + D(\E[a_{j-1}+1, a_{j}] \Phi[1,a_{j-1}](\rho)
                    \| \E[a_{j-1}+1, a_{j+1}](''))
        \\ & \geq D(\Phi[1,a_{j}](\rho) \| \E[a_{j-1}+1, a_{j}](''))
            \\ & \pl \pl \pl + D(\E[a_{j-1}+1, a_{j}] \Phi[1,a_{j}](\rho)
                    \| \E[a_{j-1}+1, a_{j+1}](''))
        \\ & = D(\Phi[1,a_{j}](\rho) \| \E[a_{j-1}+1, a_{j+1}]('')) 
        \pl ,
    \end{split}
    \end{equation}
    Combining Equations \eqref{eq:chain-range-0}, \eqref{eq:chain-range-1}, and \eqref{eq:chain-range-2} to estimate the sum in Equation \eqref{eq:induct-sequence},
    \begin{equation} \label{eq:induct-combine}
    \begin{split}
        & A_j
            + D(\Phi[1,a_{j-1}](\rho) \| \E[a_{j-1}+1, a_{j+1}](''))
        \\ & \geq 
        D(\E[a_{j-1}+1, a_{j+1}] \Phi[1, a_{j}] (\rho) \| 
            \E[a_{j-1}+1, a_{j+1}] \E[1, a_{j}] \Phi[1, a_{j}](\rho) )
        \\ & + D(\Phi[1, a_{j}](\rho) \| \E[a_{j-1} + 1, a_{j+1}](''))
        \pl.
    \end{split}
    \end{equation}
    Then using the chain rule, Equation \eqref{eq:chain-rule},
    \begin{equation} \label{eq:chain-range-combine}
    \begin{split}
        ... & \geq D(\Phi[1, a_{j}] (\rho) \| 
            \E[a_{j-1}+1, a_{j+1}] \E[1, a_{j}] \Phi[1, a_{j}](\rho) )
        \pl .
    \end{split}
    \end{equation}
    Via entropic gluing, Assumption \ref{assump:glue},
    \begin{equation}
    \begin{split}
        ... & \geq g(\E[a_{j-1}+1, a_{j+1}], \E[1, a_{j}])^{-1} D(\Phi[1, a_{j}] (\rho) \| \E[1, a_{j+1}] ('') )
        \\ & = g(\E[a_{j-1}+1, a_{j+1}], \E[1, a_{j}])^{-1} A_{j+1}
        \pl.
    \end{split}
    \end{equation}
    Combining the base inequality $B_1 \geq A_2$ with steps for $j = 2 \dots J$, the originally given sum upper-bounds
    \begin{equation}
    \begin{split}
        & \bigg ( \prod_{j=2}^J g(\E[a_{j-1}+1, a_{j+1}], \E[1, a_{j}]) \bigg )^{-1}
            D(\Phi[1, a_{J}] (\rho) \| \E[1, a_{J+1}] ('') )
         \\ &   \geq \bigg ( \prod_{j=2}^J g(\E[a_{j-1}+1, a_{j+1}], \E[1, a_{j}]) \bigg )^{-1} D(\Phi[1, m] (\rho) \| \E[1, m] ('') )
            \pl,
    \end{split}
    \end{equation}
    recalling that the $(m+1)$th channel is trivial.
\end{proof}
\begin{theorem} \label{thm:parallel-2layer-tech}
    Let $\mu$ be a unitary measure on $n$ qudits of local dimension $q$. Assume $\mu$ is induced by an $s$-local, $\ell$-layer parallel architecture admitting a $(c, \Delta)$-connected sequence with $\Delta < c$,
    \begin{equation}
        c \geq \max \{ \ell \log_q(1728 k^2), 2 \tilde{a} + 2 (1+\tilde{a}) (\ln (\tilde{a}) +
        \max \{\ln(576 k^2), \tilde{a}^2 \}) \} \pl,
    \end{equation}
    where $\tilde{a} = \max \{\ell / \ln q, 1\}$, and each subsequence or concatenated pair of adjacent subsequences has $\lambda_0$-CSDPI toward the Haar twirl on contained qudits. Then $\Phi_{\mu, k}$ is a $\lambda$-converging entropic $k$-design with $\lambda \geq \lambda_0 / 5^{\log^*(\ell n)}$. When local gates are Haar random, $s=2$, and $q=4$, $\lambda_0 = 1 / O(c \times k)$. When local gates are Haar random, and $s=2$, $\lambda_0 = 1/O(\mathrm{poly}(k))$ for fixed $q$.
\end{theorem}
\begin{proof}
    This Theorem will follow from applying Theorem \ref{thm:compose-2stream} inductively. Begin with the base case.
    By assumed $(c,\Delta)$-connectedness, the original sequence admits a partition into subsequences of length between $c$ and $c+\Delta$ (except the potentially shorter first) such that all subsequences and contiguous concatenations thereof are connected. Let $n_0 := 1 + 2(s-1)(c+\Delta)$. By Lemma \ref{lem:parallel-linear}, 2-local random circuits of size at most $2 (c + \Delta)$ have $1 / (2 \lceil (\ln q) (4 C(\sqrt{q}, k))^{\ell - 1}(4 k n_0 + \log_q(10)) \rceil)$-CSDPI, bounding $\lambda_0$, where the bounds on $C(\sqrt{q}, k)$ are noted therein. This bounds the base case or 0th level. When $s \neq 2$, an initial bound on $\lambda_0$ is assumed (not shown).
    Let $c_0 = c$ be the number of gates in such a sequence, noting that a subsequence of $r$ gates necessarily acts on at least $r / \ell$ qudits. Set $\Delta_0 = \Delta$. Let the total circuit length be $m$. If $m \leq c + \Delta$, then the Theorem reduces to the base case and is already complete.
    
    Now consider the induction step.  Assume for each $t \in \{1, \dots, \log^*m\}$ that the original subsequence is already partitioned into subsequences containing between $c_{t-1}$ and $c_{t-1} + \Delta_{t-1}$ gates (except the first, which may be shorter), and $\Delta_{t-1} < c_{t-1}$. Assume that each subsequence or pair of adjacent subsequences, when concatenated, induces a random sub-circuit with $\lambda_{t-1}$-CSDPI. Let $\exp_q(x) := q^x$. The current, level-$t$ step will concatenate these subsequences into longer subsequences, targeting a gate number of at least
    \begin{equation} \label{eq:ct}
    c_t := \lceil \exp_q(c_{t-1} / \ell - \log_q (12 c_g k^2)) \rceil
    \end{equation}
    for $c_g = 24$ as determined by Fact \ref{fact:k-design-glue}. This ensures that $c_{t-1} \leq  \ell(\log_q c_t + \log_q(12 c_g k^2))$. Via a greedy approach that, \textit{starting with the last subsequence and counting down}, simply concatenates available subsequences until reaching or exceeding the target gate number, one obtains that the level-$t$ sequences have lengths between $\lceil \exp_q(c_{t-1} / \ell - \log_q (12 c_g k^2)) \rceil$ and $\lfloor \exp_q(c_{t-1} / \ell - \log_q (12 c_g k^2)) + c_{t-1} + \Delta_{t-1} \rfloor$, except for the first, which could be shorter. Set $\Delta_t = c_{t-1} + \Delta_{t-1}$. Assuming $c_{t-1} \geq \Delta_{t-1}$, and $c_{t-1} \geq (\ell / \ln q) (\ln c_{t-1} + \ln (24 c_g k^2))$, $c_t > \Delta_t$. Lemma \ref{lem:log-compare-appended} shows that the second of these conditions is satisfied as long as
    \begin{equation} \label{eq:cond-early}
        c_{t-1} \geq (\ell / \ln q) (\ln (\ell / \ln q) + \ln(24 c_g k^2) + \ln (2 (\ln (\ell / \ln q) + \ln(24 c_g k^2)))) \pl.
    \end{equation}
    To recursively apply Theorem \ref{thm:compose-2stream} requires not just estimating CSDPI constants for the level-$t$ subsequences, but for every concatenated pair of adjacent level-$t$ subsequences. Note that since the ``$j=1$'' gluing factor in the product therein is excluded, using Theorem \ref{thm:compose-2stream} does not require lower-bounding the length of the (possibly short) first subsequence. By Theorem \ref{thm:compose-2stream}, if a concatenated pair of level-$t$ subsequence has channel $\tilde{\Phi}$ with conditional expectation $\tilde{\E}$ (including extension by an auxiliary system of arbitrary dimension), then on any input-auxiliary state $\rho$,
    \begin{equation}
    \begin{split}
        D(\rho \| \tilde{\E}(\rho)) & \geq (1 + \lambda_{t-1} (1 - c_g k^2 / q^{c_{t-1} / \ell})^{\lceil 2 (\exp_q(c_{t-1} / \ell - \log_q (12 c_g k^2)) + 2 c_{t-1})  / c_{t-1} \rceil} / 2) D(\tilde{\Phi}(\rho) \| \tilde{\E}(\rho)) \pl .
    \end{split}
    \end{equation}
    This bound also applies to single level-$t$ subsequences, as these arise from concatenating smaller numbers of blocks with the same lower-bounded overlaps. Therefore,
    \begin{equation}
        1 - \lambda_t \leq 1 / \Big ( 1 + \frac{\lambda_{t-1}}{2} \Big (1 - 1 / 3 - 12 c_g k^2 / q^{c_{t-1} / \ell} \Big ) \Big ) \pl.
    \end{equation}
    As long as $q^{c_{t-1} / \ell} \geq 72 c_g k^2$, it follows that
    \begin{equation} \label{eq:recur-lambda}
        \lambda_{t} \geq 1 - \frac{1}{1 + \lambda_{t-1}/4} \geq \lambda_{t-1}/5 \pl.
    \end{equation}
    Lemma \ref{lem:log-iterate-appended} yields for a non-decreasing function $R$ on the reals that if 
    \begin{equation}
        R(x) := \ell(\log_q x + \log_q(12 c_g k^2)) = (\ell / \ln q) (\ln x + \ln (12 c_g k^2))
    \end{equation}
    for each $j = 1 \dots \log^* m$, then
    \begin{equation}
        R^{\log^*m} (m) \leq 2 \tilde{a} + (1+\tilde{a}) (\ln \tilde{a} +
            \max \{ \ln (12 c_g k^2), \tilde{a}^2 \}) \pl.
    \end{equation}
    If $b = (\ell / \ln q) \ln(12 c_g k^2) < 1$ therein, replace it by 1 to ensure validity of the Lemma. To apply Lemma \ref{lem:iterate-inverse}, define $W(x) = \lceil \exp_q(x / \ell - \log_q (12 c_g k^2)) \rceil$. Then as long as $c_0 = c \geq R^{\log^*m}(m)$, $W^{\log^*m}(c_0) \geq m$. Hence since each elementary gate sequence (except the first) is of length at least $c$, recalling Equation \eqref{eq:ct} after $\log^*m$ levels of concatenating, $c_{\log^*m} \geq m$, and the final level sequence contains the entire sequence of gates. This constraint combines with Equation \eqref{eq:cond-early} and the aforementioned requirement that $q^{c_{t-1} / \ell} \geq 72 c_g k^2$ to require
    \begin{equation}
    \begin{split}
        c \geq & \max \{ \ell \log_q(72 c_g k^2), 2 \tilde{a} + 2 (1+\tilde{a}) (\ln (\tilde{a}) +
        \max \{\ln(576 k^2), \tilde{a}^2 \}) \} \pl.
    \end{split}
    \end{equation}
    Finally, note that $m \leq n \ell$. The Theorem then follows from iterating the recursion relation of Equation \eqref{eq:recur-lambda}.
\end{proof}
\begin{cor}[1-D Brickwork] \label{cor:brick1d-supp}
    The one-dimensional brickwork architecture with Haar-distributed 2-qubit gates is a $\lambda$-converging entropic $k$-design with
    \begin{equation}
        \lambda = \frac{1}{2 \times  5^{\log^*{(2 n)}}} \bigg \lceil \frac{(2 k (1 + 2 (c + \Delta)) + \log_2 (10))}{\log_2(1/(1-2^{-53}))} \bigg \rceil^{-1} \pl,
    \end{equation}
    whether on open or periodic (for even $n$) boundary conditions, where $c$ can be fixed according to the constraints of Theorem \ref{thm:parallel-2layer-tech}.
\end{cor}
\begin{proof}
    Illustrated in Figure \ref{fig:brickwork-order}, 1-D brickwork with open boundaries admits a gate ordering that is $(c,\Delta)$-connected for every $c \geq 2$ and $1 \leq \Delta < c$. This ordering moves from left to right (or right-to-left), at each step either applying the next second-layer gate if possible or otherwise applying the next first-layer gate. If one first applies either a first-layer gate followed by a second-layer gate one qudit to the right of it, or two first-layer gates and a second-layer gate that connects them, one obtains a connected sequence of 2 or 3 gates in which the rightmost is a first-layer gate, following a second-layer gate one qudit to its left. The next step and subsequent steps apply the next first-layer and then next second-layer gate, after which the sequence is again connected. Therefore, as long as subsequence lengths are chosen to end on second-layer gates, the subsequences are connected.
    
    Importantly, these connected subsequences of 1-D brickwork are themselves 1-D brickwork on smaller qubit ranges, as are concatenations of adjacent subsequences. Lemma \ref{lem:brickworkknown} then implies multiplicative error convergence appropriate for Equation \eqref{eq:iterate-rel} in the Supplemental Material, which yields that
    \begin{equation}
        \lambda_0 \geq \frac{1}{2} \bigg \lceil \frac{(2 k (1 + 2 (c + \Delta)) + \log_2 (10))}{\log_2(1/(1-2^{-53}))} \bigg \rceil^{-1} \pl,
    \end{equation}
    where $n_0 \leq 2 (c + \Delta) + 1$ upper-bounds the number of qubits in a concatenated subsequence pair. Substituting yields the concrete bound. One may then choose $c = \lceil c_*(k) \rceil$, where $c_*(k)$ is the lower-bound on $c$ given in Theorem \ref{thm:parallel-2layer-tech}, and apply that Theorem.
    
    Periodic boundaries reduce to open boundaries for brickwork by removing one second-layer gate, which can only reduce the CSDPI constant by the data processing inequality.
\end{proof}
\begin{theorem} \label{thm:random-graph-tech}
Assume $G$ is a connected, $n$-node graph of degree at most $c$. Then $\Phi_{G,k}$ is a $\lambda_0^G/(c 64^{\log^*n} n)$-converging entropic $k$-design, where $\lambda_0^G$ is the minimum CSDPI constant for every connected, max-degree $c$ subtree of size between $2$ and
\begin{equation}
\begin{split}
    & n_0 = \max \{6, 2 \tilde{a} + (1 + \tilde{a}) (\ln (\tilde{a})
        + \max\{\tilde{a}^2, \ln (96 k^2) + 2 \} ) \} \pl,
\end{split}
\end{equation}
of a fixed spanning tree, where $\tilde{a} = \max \{ 2(c+1) / \ln q, 1\}$. Let $\lambda_{\mathrm{edge}}$ denote the minimum CSDPI constant of any edge's random gate toward the local Haar twirl, which is 1 for local Haar twirls. In general, $\lambda_0^G = \lambda_{\mathrm{edge}} / O(\mathrm{poly}(k))$, ignoring $q$ and $c$ dependence. When $q=4$, $\lambda_0^G = \lambda_{\mathrm{edge}} / O(k (\log k)^2)$ ignoring $c$ dependence. When $q=2$ for a cycle or path graph,
\begin{equation}
    \lambda_{0}^G \geq \lambda_{\mathrm{edge}} / ( 2 n_0 c \lceil (2 n_0 k + \log_2(24)) / \log_2(1/(1-2^{-53})) \rceil) \pl.
\end{equation}
\end{theorem}
\begin{proof}
    For a given graph $G'$, let $S$ denote a given family of subgraphs of $G'$. Let $E(G')$ denote the edge set of $G'$, and $|E(G')|$ denote the number of edges in $G'$. Let $M_S(G')$ be defined as maximum over each in $E(G')$ of the number of subgraphs in $S$ that contain $e$ as an edge.
    Let
    \begin{equation}
        p := \frac{1}{M_S(G')}  \sum_{H \in S} \frac{|E(H)|}{|E(G')|} \pl,
    \end{equation}
    and
    \begin{equation}
        \Theta := \frac{1}{p} \sum_{H \in S} \frac{|E(H)|}{|E(G')| M_S(G')} \Phi_{H, k} \pl.
    \end{equation}
    We may rewrite
    \begin{equation} \label{eq:decomp-graph}
        \Phi_{G', k} = p \Theta +  (1-p) \Psi
    \end{equation}
    for a remainder channel $\Psi$. By convexity of relative entropy, $\Phi_{G', k}$ has CSDPI constant at least $p$ times that of $\Theta$.

    If $G$ admits a spanning tree with fewer than $n_0$ vertices, then Equation \eqref{eq:tree-graph} yields the desired estimate. The rest of the proof will assume otherwise.

    Let $T$ be a tree of maximum degree $c$. Set a to-be-determined $r \in \NN$. Assume as an induction hypothesis that for every tree $T'$ with at most $2 r$ nodes and degree at most $c$, $\Phi_{T', k}$ has $\lambda'$-CSDPI.
    $T$ necessarily admits a walk $W$ given by a depth-first traversal that visits every node at least once and at most $c+1$ times, and every edge at most twice. Therefore, dividing $W$ into subsequences $W = W_1 + \dots + W_v$ for $v \in \NN$ (``$+$'' denotes concatenation), any subsequence $W_i$ of length $|W_i|$ (counting steps, or 1 plus the number of edges) visits at least $|W_i| / (c+1)$ and at most $|W_i|$ nodes. Each node is visited at most $c+1$ times. Let $W_1, \dots, W_v$ be defined so that each $W_i$ has length $r$, except for the last, which might be smaller and does not affect gluing constants. For each pair $i = 1 \dots v-1$, let $H[i, i+1]$ be the sub-tree of $G$ including each of the edges used in $W_i + W_{i+1}$. Hence, the first half of each $H[i,i+1]$ arising from the $W_i$ component includes at least $r / (c+1)$ nodes. The total of each $H[i,i+1]$ contains at least $r / (c+1)$ nodes, at least $(r-1) / 2$ edges, at most $2 r$ nodes, and no more than $2 r$ edges. Let $\E[i,i+1]$ denote the fixed point conditional expectation of $\Phi_{H[i,i+1], k}$ for each $i = 1 \dots v-1$. Note that $v - 2 \leq 2 |E(T)|/(r-1)$ using the depth-first traversal length.
    
    By Fact \ref{fact:k-design-glue}, the family $(\E[i,i+1])_{i=1}^{v-1}$ quasi-factorizes with each $\alpha_i \leq (1 - c_g k^2 / q^{r/(c+1)})^{-(v-2)}$ for $c_g = 24$. Apply Equation \eqref{eq:decomp-graph} with $M_S(G') \leq 4$, $|E(H[i,i+1])| \geq (r-1) / 2$, and $G' = T$, using Lemma \ref{lem:decay-merge} to estimate the CSDPI constant of the first sum therein. Hence, the CSDPI constant $\lambda_T$ of $\Phi_{T, k}$ is at least $\lambda' \times (1 - c_g (v-2) k^2 / q^{r / (c+1)}) \min_i |E(H[i,i+1])| / (4 |E(T)|)$. So that $(1 - 2 c_g (v-2) k^2 / q^{r/(c+1)}) \geq 1/2$, set $r = \max \{3, \lceil (c+1) \log_q (4 c_g |V(T)| k^2) \rceil \}$. Observe that
    \begin{equation}
        v - 2 \leq \frac{2 |E(T)|}{r-1} \leq |V(T)|
    \end{equation}
    when $r \geq 3$, yielding together with $q^{r/(c+1)} \geq 4 c_g |V(T)| k^2 $ that
    \begin{equation}
        \lambda_T \geq \frac{r-1}{16|E(T)|} \lambda' 
    \end{equation}
    
    Next, we will recursively decompose the original tree as above. Let $j$ denote the $j$th level of recursion, and $t_j$ upper bound the largest possible tree size (in nodes) at the $j$th level.
    Let $t_0 := |V(T)|$, and $T_0 = T$. Let $\lambda(T) := \sup \{\lambda : \Phi_{T,k} \text{has $\lambda$-CSDPI} \}$, and $\lambda_j$ be the smallest $\lambda(T')$ for any subtree $T' \subseteq T_0$ such that $2 \leq |V(T')| \leq t_j$for each $j \geq 1$. Then let $\lambda_0 := \lambda(T_0)$ . The tree size estimates then follow the recursion relation
    \begin{equation} \label{eq:tree-size-recur}
        t_{j+1} = 2 r_j = 2 \max \{ 3, \lceil (c+1) \log_q ( 4 c_g t_j k^2) \rceil \} \pl,
    \end{equation}
    where $r_j$ is the per-layer subsequence length, and at each step the trees shrink. Note that for every $t \geq n_0$,
    \begin{equation}
        t > 2 \max \{ 3, \lceil (c+1) \log_q ( 4 c_g t k^2) \rceil \} \pl,
    \end{equation}
    so the trees indeed shrink layer-by-layer. If $t_{j+1} \leq 6 \leq n_0$, then the recursion stops. Otherwise, for each subtree $T$ at the $j$th level and corresponding walk $W$ \dots
    \begin{itemize}
        \item If $|V(T)| \leq t_{j+1}$, use $\lambda_{T} \geq \lambda_{j+1}$, treating $T$ as equivalent to a $(j+1)$th level tree.
        \item otherwise decompose via walks of lengths all equal to $r_j$ except possibly for the last.
    \end{itemize}
    Since Equation \eqref{eq:tree-size-recur} implies that $t_{j+1} \leq t_j$, $(t_{j+1} - 2) / 32 t_j 
    \leq 1$. Then because $|E(T)| \leq t_j$, in both cases
    \begin{equation} \label{eq:tree-rate-recur}
        \lambda_T \geq \frac{r_j - 1}{16 t_j} \lambda_{j+1} \geq \frac{t_{j+1} - 2}{32 t_j} \lambda_{j+1} \pl.
    \end{equation}

    Apply Lemma \ref{lem:log-iterate-appended} with $a = 2 (c+1) / \ln q$ and $b \leq a \ln (4 c_g k^2) + 2$. To simplify possible outcomes, note that we are always free to replace $a$ by a larger constant, so we may assume $\ln q \geq 1/2$. Then
    \begin{equation} \label{eq:max-x-graph}
    \begin{split}
        t_{\log^*(t_0)} \leq &
             \max \{6, 2 \tilde{a} + (1 + \tilde{a}) (\ln (\tilde{a})
        + \max\{\tilde{a}^2, \ln (4 c_g k^2) + 2 \}) \} \pl.
    \end{split}
    \end{equation}
    Let $x$ be the first index for which $t_x \leq n_0$, where $x \leq \log^* t_0 \leq \log^* n$. Therefore, the Theorem's conditions ensure that $\lambda_x \geq \lambda_0^G$. After $x$ levels of induction, Equations \eqref{eq:tree-size-recur} and \eqref{eq:tree-rate-recur} (substituting $t_j$ for $|E(T)|$ and noting that $|E(T)| \leq |V(T)|$) imply that
    \begin{equation}
        \lambda_0 = \lambda_T \geq \lambda_{x} \prod_{j=0}^{x-1} \frac{t_{j+1}-2}{32 t_j}
        \geq \lambda_{x} \prod_{j=0}^{x-1} \frac{t_{j+1}}{64 t_j}
        = \left ( \frac{1}{64} \right )^x \frac{t_{x}}{t_0} \lambda_{x}
        \geq \left ( \frac{1}{64} \right )^x \frac{1}{t_0} \lambda_{x} \pl,
    \end{equation}
    assuming that $t_{x} \geq 4$, which is already satisfied by the recursion stopping at or above a minimum size of 6.
    
    To complete the general bounds, note that
    \begin{equation} \label{eq:tree-graph}
        \Phi_{G, k} = (|E(T)|/|E(G)|) \Phi_{T,k} + (1 - |E(T)|/|E(G)|) \Phi_{G - T,k}
    \end{equation}
    for some spanning tree $T$, where $G - T$ is the graph with all edges in $T$ removed. Because the number of edges in $T$ is at least the number of nodes in $G$ minus one, $|E(T)|/|E(G)| \geq 2(n-1)/c n \geq 1/c$. Finally, to control the constant $\lambda_0^G$ in known cases, apply Lemma \ref{lem:small-random}.

    For the special case of a cycle or path graph, Lemma \ref{lem:small-random} yields a specific bound on $\lambda_0^G$. The path graph is already a tree, so $T$ can be taken as the original graph itself. A cycle reduces to a path by removing and ignoring one edge, and to avoid slightly reducing the constant, one may take the cycle to be the average over the paths yielded by removing each edge.
\end{proof}

\begin{figure}[ht!] \centering
	\begin{subfigure}[b]{0.35\textwidth}
		\includegraphics[width=0.80\textwidth]{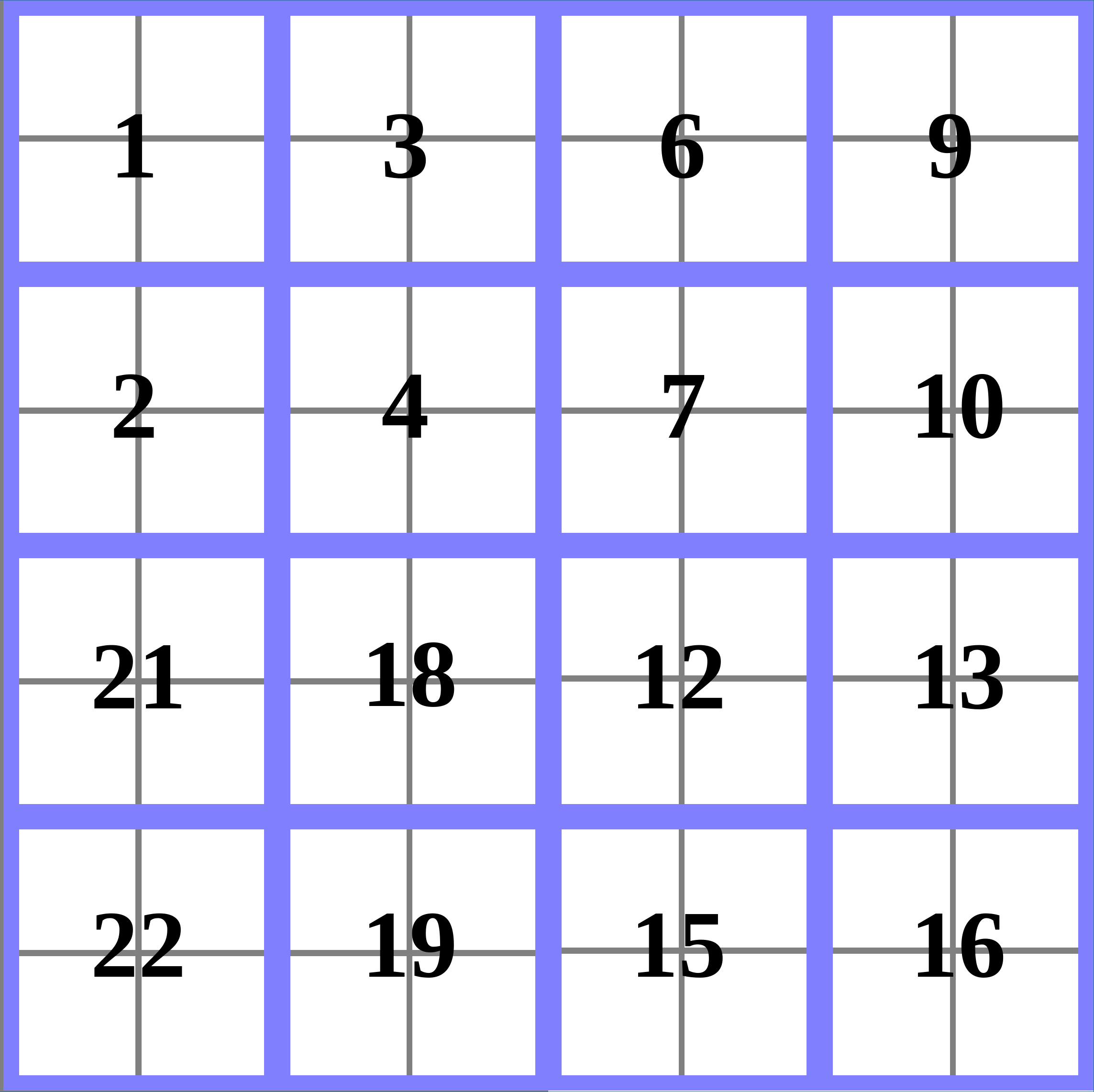}
        \caption{}
	\end{subfigure}
	\hspace{5mm}
	\begin{subfigure}[b]{0.35\textwidth}
		\includegraphics[width=0.80\textwidth]{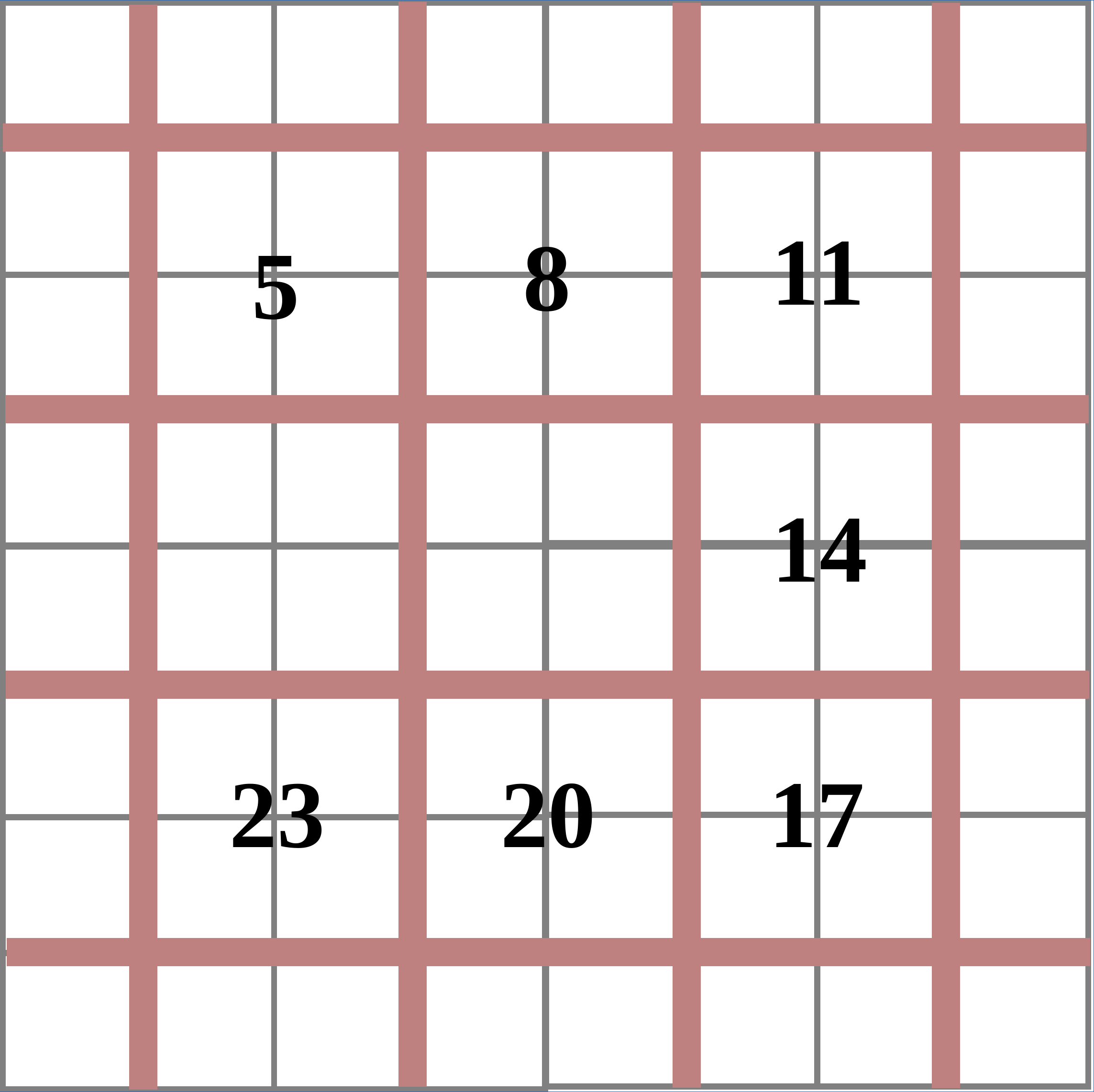}
        \caption{}
	\end{subfigure}
    \caption{Illustration of a 2-D square lattice architecture forming a shallow entropic $k$-design. Each small square corresponds to a qudit, and each cluster of 4 as delimited by thick, colored lines defines the application of a random gate. The numbers define a path through the connectivity hypergraph, each number labeling an edge by its place in the path. (A) represents the first layer and (B) the second. Due to periodic boundary conditions, the rightmost and bottom clusters in (B) extend respectively to the left and top.}
    \label{fig:lattice2D}
\end{figure}
\subsection{Conditional Extension to Higher-Dimensional Parallel Grids} \label{sec:lattice}
The applicability of Theorem \ref{thm:parallel-2layer} extends beyond 1-D brickwork to hypercubic lattices with periodic boundaries made of unit hypercubes, as illustrated for 2-dimensional spatial lattices in Figure \ref{fig:lattice2D}. Consider a $D$-dimensional hypercube lattice formed from $n = x^D$ qubits with side length $x$ and periodic boundary conditions. Consider the protocol as in Figure \ref{fig:lattice2D} that applies random gates to adjacent $(D-1)$-dimensional, length-2 hypercubes in the first (temporal) layer. Each time a new $(D-1)$-dimensional hypercube is created (except for the first), a second-layer gate is applied to `glue' it to the previous. After the first layer is completely finished, the rest of the second layer can be applied via post-processing. This gate ordering forms a  $(2^{D-1}, 2)$-connected sequence. These may be grouped into longer subsequences to satisfy the conditions of Theorem \ref{thm:parallel-2layer}.
Therefore, in each spatial dimension $D$, if the architecture induced by any contiguous subregion $R$ converges to an $\epsilon$-error multiplicative design in depth $\tilde{C}(k) (n_R k + \log (1/\epsilon))$, where $n_R$ is the number of qudits in the subregion, then for the induced measure $\mu$, $\Phi_{\mu, k}$ has $\lambda_0 / 5^{\log^* (2 n)}$-CSDPI with $D$-dependent constants. Note that this scenario lacks a formal, pre-existing estimate of $\lambda_0$. It is reasonable to conjecture that $\lambda_0 = 1/O(k \mathrm{polylog}(k))$ for fixed $s$ and $q$, although it remains open to precisely determine.

\end{document}